\documentclass[conference,compsoc]{IEEEtran}
\IEEEoverridecommandlockouts
\usepackage[nocompress]{cite}
\usepackage{amsmath,amssymb,amsfonts}
\usepackage{algorithmic}
\usepackage{graphicx}
\usepackage{textcomp}
\usepackage{xcolor}
\usepackage{subcaption}
\usepackage{comment}
\usepackage{booktabs}
\usepackage{multirow}
\usepackage{makecell}
\usepackage[most]{tcolorbox}
\usepackage{enumitem}
\usepackage{ragged2e}
\usepackage{placeins}
\usepackage{float}
\usepackage{array}
\usepackage{tabularx}
\usepackage{marvosym}
\usepackage{adjustbox}

\usepackage[hidelinks]{hyperref}
\newcommand{\cmark}{\ding{51}}
\newcommand{\xmark}{\ding{55}}
\usepackage{siunitx}
\usepackage{listings}
\usepackage{calc}
\usepackage{xspace}
\usepackage{pifont}

\newcommand{\ignore}[1]{}

\newcommand{\name}{\textsc{Lens}\xspace}

\usepackage{xcolor}
\usepackage[normalem]{ulem}

\newcommand{\magentaadd}[1]{{#1}}

\makeatother

\makeatletter
\def\ps@accepted{%
  \def\@oddhead{\hfil\raisebox{5mm}[0pt][0pt]{\normalfont\footnotesize
    Accepted at the 2027 IEEE Symposium on Security and Privacy (S\&P 2027).}\hfil}%
  \let\@evenhead\@oddhead
  \def\@oddfoot{}%
  \let\@evenfoot\@oddfoot
}
\makeatother

\begin{document}

\title{When Core-Network Elements Become Reachable:\\Field-Level Exploitability Analysis in Diameter}

\author{

\IEEEauthorblockN{
Wei Yi$^{1,2}$, Changlong Han$^1$, Xue Zhang$^2$, Qiao Zhang$^2$,
 \\Yi Chen$^{2,\text{\Letter}}$, Jin Cao$^{1,\text{\Letter}}$, Di Tang$^3$,  Yulong Fu$^1$, Ben Niu$^4$, Hui Li$^1$
}

\IEEEauthorblockA{
$^1$Xidian University 
$^2$The University of Hong Kong 
$^3$Sun Yat-Sen University 
$^4$Institute of Information Engineering,CAS
}

\IEEEauthorblockA{
\{yiwei@stu., clhan\_1@stu., jcao@, ylfu@, lihui@mail.\}xidian.edu.cn  \\
\{chenyi1@, xue.zhang@connect., qiaozhang@connect.\}hku.hk \quad tangd9@mail.sysu.edu.cn \quad 
niuben@iie.ac.cn
}
\thanks{$^{\text{\Letter}}$Corresponding authors.}
}

\vspace{-5pt}
\maketitle
\thispagestyle{accepted}
\vspace{-5pt}
\begin{abstract}

The increasing exposure of cellular core-network elements to public networks has turned Diameter signaling interfaces into a practical attack surface. While modern large language models (LLMs) show promise for protocol exploitability analysis, we observe that their judgments on Diameter often become unreliable, as they may deviate from the tacit analysis discipline followed by security experts and prematurely reach conclusions without considering the complete protocol evidence needed for sound analysis.
To address this challenge, we present \name, an LLM-based system for reliable field-level Diameter exploitability analysis. Rather than directly asking LLMs whether a Diameter field is exploitable, \name guides LLMs using both the exploitability-analysis discipline and the protocol evidence needed for exploitability analysis. To obtain such guidance, \name leverages LLMs to infer analysis discipline from existing exploit writeups and evidence-collection guidance from exploitability-analysis failures. The inferred guidance enables \name to systematically collect protocol evidence, assess the exploitability of target Diameter fields, and generate proof-of-concept (PoC) attacks when applicable.

\magentaadd{Our experiments show that \name performs reliable exploitability analysis across four LLMs and repeated runs, achieving mean recall of 94.00\% to 98.00\%, precision of 78.73\% to 88.94\%, and accuracy of 84.00\% to 92.00\%.}
Using \name, we find that exploitability is surprisingly widespread in Diameter: more than 25\% of field paths in the S6a and Cx interfaces are exploitable under the considered threat model. Building on these protocol-level findings, we further assess their impact on practical implementations and uncover 95 previously unknown vulnerabilities, including 17 in open-source EPC/IMS implementations and 78 in commercial deployments. These vulnerabilities lead to severe security consequences, such as service disruption, information leakage, subscriber tracking, and accelerated battery depletion. Furthermore, our results demonstrate the strong generalizability of \name. Applying \name to 16 additional protocols spanning cellular, network, industrial, and host-internal systems uncovers 118 previously unknown vulnerabilities.

\end{abstract}
\vspace{-5pt}
\section{Introduction}
\label{sec: Introduction} 
\vspace{-5pt}

Cellular core networks have long been treated as trusted internal environments. Recent research and security reports~\cite{zhang2025invade, miller2026badconnection,yang2026smallcell}, however, show that this assumption does not always hold in practice. 4G core-network elements have increasingly been found reachable from public networks, and existing works further demonstrate that attackers can exploit such exposure through crafted signaling messages, potentially triggering unintended core-network operations and security consequences.
While existing works provide only isolated attack cases~\cite{miller2026badconnection}, public exposure of core-network elements is now a practical reality. As a result, systematically understanding the broader security risks introduced by exposed core-network signaling interfaces becomes increasingly important.

Among core-network signaling protocols, the Diameter protocol is a particularly security-critical target~\cite{rfc6733,29230}. Diameter is used by core-network elements to exchange and update subscriber information, supporting critical network functions such as subscriber authentication, service access, and billing. Crafting Diameter messages with manipulated field values may therefore cause core-network elements to make incorrect security-sensitive decisions, potentially affecting subscriber services and triggering unintended core-network behavior.
However, systematically determining whether a Diameter field is actually exploitable is highly challenging. A Diameter field that appears harmless within its local message context may later influence security-sensitive behavior after its value is processed and propagated across multiple protocol procedures and core-network elements. In many cases, understanding such effects requires correlating protocol logic scattered across multiple and complex 3GPP specifications, making exploitability analysis difficult even for domain experts. Moreover, Diameter messages contain hundreds of fields. Systematically analyzing their exploitability manually is therefore impractical, and automated analysis becomes necessary.

Given the capabilities of modern large language models (LLMs), a natural question is whether they are already sufficient for this task. Encouragingly, our preliminary study shows that even directly asking LLMs whether a Diameter field is exploitable already works surprisingly well in some cases: LLMs can not only correctly judge whether a Diameter field is exploitable, but also generate working proof-of-concept (PoC) exploits. This suggests that modern LLMs are already capable of performing non-trivial exploitability analysis for Diameter protocols.
However, such capability remains highly unstable in practice. The direct use of LLMs still fails on more than half of the cases in our preliminary experiments. Upon closer examination, we find that these incorrect judgments mainly arise from two issues.
First, LLMs may drift beyond the tacit analysis discipline commonly followed by security experts during exploitability analysis. For example, LLMs may assume attacker capabilities that experts would normally rule out in practice, such as assuming that attackers can obtain long-term shared secrets between trusted core-network components. Such drift can easily lead to incorrect exploitability conclusions.
Second, LLMs may prematurely make judgments without considering all protocol evidence relevant to exploitation. For example, LLMs may overlook how crafted field values later influence subsequent core-network operations, even when such information already exists in 3GPP specifications. Without considering sufficient protocol evidence, LLMs can easily make incorrect judgments.

These observations suggest a promising direction for using LLMs to perform reliable Diameter exploitability analysis: explicitly providing LLMs with the \textit{tacit exploitability-analysis discipline} used by security experts and the \textit{protocol evidence} relevant to exploitation.
However, obtaining both remains highly challenging.
For tacit analysis discipline, it is nearly impossible to force experts to fully write it out. This challenge resembles Polanyi’s paradox~\cite{polanyi2009tacit}: \textit{``we know more than we can tell,''} conveying that fully articulating tacit expert knowledge into explicit rules is difficult for humans.
For protocol evidence, even experts usually do not know all relevant information upfront. In practice, exploitability analysis is typically a failure-driven process with iterative evidence seeking: analysts continuously search for additional protocol information from 3GPP specifications to understand why current exploit attempts fail, or determine whether a target field is exploitable or not.

\vspace{2pt}\noindent\textbf{Our method}.
Facing these challenges, our method leverages LLMs to infer both tacit analysis discipline and protocol evidence for Diameter exploitability analysis.
For tacit analysis discipline, given LLMs' strong abstraction and induction capabilities~\cite{honovich2023instruction}, we ask LLMs to infer the tacit analysis discipline from existing exploit writeups.
This is possible because exploit writeups naturally reflect such discipline: the exploitability analysis and PoC construction described in these writeups are produced under the discipline's guidance.
For protocol evidence, given LLMs' reflective reasoning capabilities~\cite{cai2024large, lin2025ado}, we let LLMs emulate how human experts infer missing protocol evidence from failed exploit attempts.
Specifically, we ask LLMs to analyze inconsistencies between incorrect exploit writeups generated by LLMs and corresponding ground-truth ones, and further infer what protocol evidence should be considered for LLMs to make correct exploitability judgments.
Here, rather than asking LLMs to infer the exact missing protocol evidence for a specific case, our method asks LLMs to infer general guidance for collecting protocol evidence, which can be applied to any target field.

Building on this design, we propose \name, a carefully designed LLM-based system for reliable fie\textbf{L}d-level Diamet\textbf{E}r exploitability a\textbf{N}aly\textbf{S}is.
Given a target Diameter field, \name first uses the LLM-inferred evidence-collection guidance to collect protocol evidence associated with the target field.
It then provides the collected evidence, together with the inferred tacit analysis discipline, as guidance for LLMs to perform exploitability analysis.
When the target field is determined to be exploitable, \name further generates corresponding PoCs. 
These PoCs are also used for security-impact assessment of representative core-network implementations in our controlled testing environment.

\vspace{2pt}
\noindent\textbf{Results.}
\magentaadd{We evaluate \name\ on a manually labeled dataset of Diameter fields and find that it performs reliable exploitability analysis across four LLMs and repeated runs, achieving mean recall of 94.00\% to 98.00\%, mean precision of 78.73\% to 88.94\%, and mean accuracy of 84.00\% to 92.00\%.}
Using \name, we then perform the first systematic field-level exploitability analysis of the Diameter protocol.
\magentaadd{The results reveal widespread field-level exploitability: among 766 analyzed fields in the S6a and Cx interfaces, 204 fields (26.63\%) are exploitable at the protocol level under our threat model, indicating that over one quarter of the analyzed Diameter fields can be exploited once core-network elements become reachable.}
Remarkably, \magentaadd{196} of them have not been previously reported. 
These weaknesses are not merely protocol-level findings. Through implementation-level assessment on representative EPC/IMS implementations, we uncover 95 previously unknown vulnerabilities, including 17 in open-source implementations and 78 in commercial implementations. These vulnerabilities enable severe security consequences affecting real subscribers, including persistent service disruption, information leakage, subscriber tracking, and accelerated battery depletion. 
Finally, we demonstrate that the methodology behind \name\ generalizes beyond Diameter.
Using \name\ to analyze 16 additional widely used protocols spanning diverse domains, such as VoIP communication, Internet routing, and IoT messaging, uncovers 279 protocol-level PoC\magentaadd{\ candidates}, including 150 previously unreported ones, of which 118 are further confirmed in real open-source implementations.

\vspace{2pt}\noindent\textbf{Contribution}. This paper makes the following contributions:

\vspace{1pt}\noindent$\bullet$~\textit{New technique.} 
We present \name, an LLM-based methodology that infers exploitability-analysis discipline and evidence-collection guidance for reliable field-level protocol exploitability analysis.

\vspace{1pt}\noindent$\bullet$~\textit{New finding.}
We conduct the first systematic field-level exploitability analysis of the Diameter S6a and Cx interfaces, revealing that \magentaadd{over one quarter} of the analyzed fields are exploitable once core-network elements become reachable and uncovering 95 previously unknown vulnerabilities in open-source and commercial EPC/IMS implementations.

\vspace{-5pt}
\section{Background}
\label{sec: Background}
\vspace{-5pt}
\begin{figure*}
    \centering
    \includegraphics[width=1\textwidth]{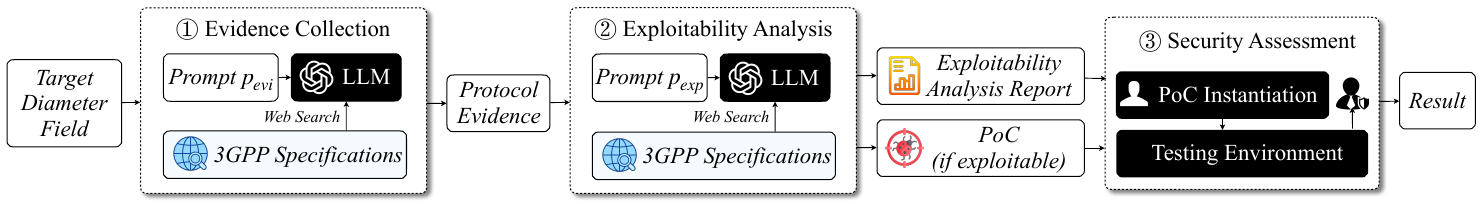}
    \caption{Workflow of \name.}
    \label{fig: Workflow of LENS}
    \vspace{-10pt}
\end{figure*}

\vspace{-5pt}
\noindent\textbf{Diameter protocol}.
Diameter is a signaling protocol family widely used in 4G core networks to support communication among network elements (NEs)\cite{29230}. Different Diameter-based interfaces are responsible for different network functions. For example, the S6a interface uses Diameter signaling between the Mobility Management Entity (MME) and the Home Subscriber Server (HSS) to support subscriber management and authentication, while the Cx interface uses Diameter signaling between IP Multimedia Subsystem (IMS) entities and the HSS for IMS service control (e.g., voice call services).
Through these interfaces, NEs use Diameter messages to request information, deliver instructions, or make control-plane decisions. In particular, the semantics of Diameter messages are determined by the values of message fields defined in 3GPP technical specifications. Given a specific interface and message type, setting individual or combined field values instructs NEs to perform different operations.
In this work, we focus on the S6a~\cite{29272} and Cx~\cite{29228,29229} interfaces, as they are representative Diameter-based interfaces in 4G core networks.

\vspace{2pt}\noindent\textbf{Threat model}.
\label{subsec:threat}
We consider the following threat model: a victim NE exposes a Diameter interface reachable from public networks and does not enable TLS protection for incoming Diameter connections; an attacker can impersonate a legitimate NE, establish Diameter connections with exposed NEs, and send crafted Diameter messages with manipulated field values to influence subsequent network operations.
This threat model actually exists in real-world deployments, according to recent research and security reports~\cite{zhang2025invade,cui2025security,yang2026smallcell, miller2026badconnection, kwon2021towards}. In addition, through our communications, mobile network operators and core-network equipment vendors further confirmed that exposed internal NEs and disabled TLS protection are not uncommon in commercial environments due to maintenance and performance considerations.\footnote{We omit the identities of the involved operators and vendors for confidentiality reasons.}

\vspace{-5pt}
\section{Design}
\label{sec: Design}

\vspace{-5pt}
\subsection{Overview}
\label{subsec: Overview}
\vspace{-5pt}

\name aims to enable reliable LLM-based field-level Diameter exploitability analysis. To achieve this goal, \name is designed to provide LLMs with two types of information, which are often overlooked when directly asked whether a Diameter field is exploitable: (1) the \textit{tacit exploitability-analysis discipline} used by security experts and (2) the \textit{protocol evidence} needed for exploitability analysis and PoC construction.

However, as discussed in the Introduction, obtaining such information is challenging, even for security experts. To address this challenge, our design leverages LLMs to derive such information. For exploitability-analysis discipline, we ask LLMs to infer the discipline from existing exploit writeups. For protocol evidence, we ask LLMs to analyze the inconsistencies between incorrect LLM-generated exploit writeups and the corresponding ground-truth ones, and then infer how protocol evidence should be collected. Both inferred results are encoded into two prompts, $p_{exp}$ and $p_{evi}$: $p_{exp}$ encodes exploitability-analysis discipline, and $p_{evi}$ encodes evidence-collection guidance. Notably, the two prompts are not independent, as inferring either prompt requires information provided by the other. \name therefore refines the two prompts jointly.

Once the two prompts are prepared, \name uses them to analyze previously unanalyzed Diameter fields. Given a target Diameter field, \name first performs evidence collection using $p_{evi}$ to gather protocol evidence associated with the target field across the 3GPP specification corpus. The collected evidence is then provided together with the target field to $p_{exp}$ for exploitability analysis, which generates an exploitability analysis report and a corresponding PoC when the target field is determined to be exploitable. Finally, the generated PoCs are used by \name for security-impact assessment. Specifically, the generated PoCs are executed against representative core-network implementations in our controlled testing environment to examine whether these implementations are affected in the manner predicted in the exploitability analysis report. Fig.~\ref{fig: Workflow of LENS} illustrates the overall workflow of \name.

The remainder of this section details how \name infers exploitability-analysis discipline and evidence-collection guidance, jointly refines the resulting prompts, and performs security-impact assessment using the generated PoCs.

\vspace{-6pt}
\subsection{Exploitability-Analysis Discipline Inference}
\label{subsec: p_exp infer}
\vspace{-6pt}

Diameter exploitability analysis is typically performed under a tacit analysis discipline used by security experts. It guides how exploitability conclusions are reached and how PoCs are constructed. However, as discussed in the Introduction, this discipline is tacit and difficult for human experts to fully write out as a set of rules.
Our key idea is to infer this discipline from existing exploit writeups. This design is based on the observation that exploit writeups naturally reflect how security experts apply the exploitability-analysis discipline in practice. Typically, a writeup includes the threat model, which specifies attack assumptions and attacker capabilities, the exploitability analysis explaining why the target field is considered exploitable or non-exploitable, and, when applicable, the attack procedure and corresponding PoCs. In this way, exploit writeups naturally preserve useful traces of the discipline. Therefore, \name leverages existing exploit writeups to infer the exploitability-analysis discipline and subsequently uses it to guide Diameter exploitability analysis.

\begin{figure}
    \centering
    \includegraphics[width=1\linewidth]{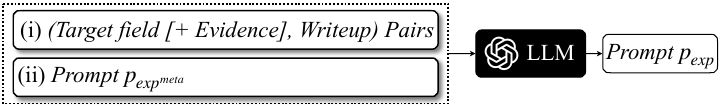}
    \caption{Inference of the exploitability-analysis prompt $p_{exp}$.}
    \label{fig: p_exp}
    \vspace{-5pt}
\end{figure}

To this end, \name leverages an LLM to infer the discipline and represent it as a prompt, denoted as $p_{exp}$. This representation is natural because the exploitability-analysis discipline itself is essentially an instruction describing how exploitability analysis should be performed. Consequently, representing it as a prompt allows it to be directly provided to an LLM, enabling the inferred discipline to guide subsequent Diameter exploitability analysis, as shown in Step \ding{193} of Fig.~\ref{fig: Workflow of LENS}.
Fig.~\ref{fig: p_exp} illustrates the process used by \name\ to infer $p_{exp}$. Specifically, we first construct a set of pairs from existing expert-written exploitability-analysis writeups. Each pair consists of (i) a target Diameter field together with its supporting protocol evidence and (ii) the corresponding expert-written writeup. Depending on the exploitability of the target field, the writeup may either document a successful exploitation case, including the corresponding attack procedure and PoC, or explain why exploitability cannot be established under the assumed threat model.
These pairs define the desired inputs and outputs of exploitability analysis. By organizing expert-written analyses into such pairs, we can easily provide the LLM with examples of the desired analysis task. Once such pairs are constructed, we provide them to an LLM with a simple meta-prompt, denoted as $p_{{exp}^{meta}}$ (shown in Table~\ref{table: p_exp_meta}), which asks the LLM to infer a prompt that can derive the latter from the former and apply the same analysis process to new Diameter fields. 

This design is inspired by recent studies showing that, given examples of task inputs and their corresponding outputs, LLMs can infer task instructions that reproduce the input-output mapping and generalize to unseen cases~\cite{honovich2023instruction}. Following the same intuition, \name\ leverages an LLM to infer $p_{exp}$ from expert-written exploitability-analysis examples. 
Interestingly, our experiments suggest that the exploitability-analysis discipline can be inferred from surprisingly few examples. In particular, one exploitable and one non-exploitable example are already sufficient in our setting. Therefore, \name\ uses a randomly selected exploitable example and a randomly selected non-exploitable example as seed examples for inferring $p_{exp}$. Detailed experimental results are presented in Section~\ref{subsec: configuration study}.

\vspace{-6pt}
\subsection{Evidence-Collection Guidance Inference}
\label{subsec: p_evi infer}
\vspace{-6pt}

As discussed in the Introduction, protocol evidence is essential for Diameter exploitability analysis, yet remains difficult to collect in advance, even for security experts. In practice, experts typically follow a failure-driven process: they first perform analysis using currently available information and, when they find such information insufficient to support a conclusion, they revisit the relevant protocol specifications to identify and obtain the missing protocol evidence needed for further analysis.
Recent studies have shown that LLMs can also learn from task failures and use failure feedback to infer how future inputs should be improved~\cite{cai2024large,lin2025ado,xu2026beyond,wang2025llms,yu2024auto}. This capability naturally aligns with the failure-driven process used by experts to collect protocol evidence. Therefore, \name\ leverages this capability to infer general evidence-collection guidance from exploitability-analysis failures.

\begin{figure*}
    \centering
    \includegraphics[width=0.9\textwidth]{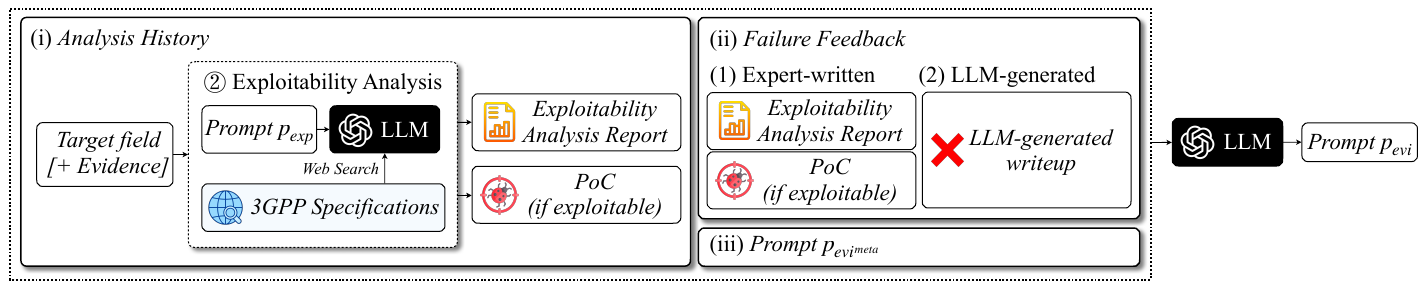}
    \vspace{-5pt}
    \caption{Inference of the evidence-collection prompt $p_{evi}$.}
    \label{fig: p_evi}
    \vspace{-7pt}
\end{figure*}

\begin{table}[t]
\centering
\caption{Template of $p_{\mathrm{exp}}^{\mathrm{meta}}$.}
\label{table: p_exp_meta}
\vspace{-3pt}

\begingroup
\footnotesize
\normalfont
\setlength{\tabcolsep}{0pt}
\renewcommand{\arraystretch}{1}

\newcommand{\textindent}{0.3em}
\newcommand{\exampleindent}{1.2em}
\newcommand{\ioindent}{1em}

\begin{tabular}{@{}>{\raggedright\arraybackslash}p{\columnwidth}@{}}
\toprule

{\normalfont\fontfamily{fvm}\selectfont
{\leftskip=\textindent
\noindent You are given several input-output examples for Diameter field analysis.
Each input describes a target Diameter field together with its protocol evidence. Each output is an expert-written writeup, including an exploitability analysis report and a PoC (if the field is exploitable).
\par\medskip
\noindent Your task is to infer a prompt that captures the exploitability-analysis discipline reflected by these examples and can guide analysis of unseen Diameter fields. Output only the inferred prompt.
\par}
}

\par\smallskip

{\normalfont\fontfamily{fvm}\selectfont
{\leftskip=\exampleindent
\noindent\textbf{Example format:}\par\vspace{0.2em}
\noindent\hspace*{\ioindent}%
\begin{tabular}{@{}>{\normalfont\footnotesize\fontfamily{fvm}\selectfont}l@{}}
Input: \{\texttt{target field and protocol evidence}\} \\
Output: \{\texttt{exploitability-analysis writeup}\}
\end{tabular}
\par}
}

\\
\bottomrule
\end{tabular}
\endgroup
\vspace{-7pt}
\end{table}

Fig.~\ref{fig: p_evi} illustrates how \name\ infers evidence-collection guidance. Specifically, when exploitability analysis performed using $p_{exp}$ did not pass expert review, \name\ provides an LLM with three inputs: (i) an analysis history, (ii) a failure feedback, and (iii) a meta-prompt ($p_{{evi}^{meta}}$), and asks the LLM to infer the evidence-collection guidance, represented as a prompt ($p_{evi}$).
First, the analysis history records a previous LLM-based exploitability-analysis attempt: given a target Diameter field, the available protocol evidence (if any), and the prompt $p_{exp}$, the LLM generated a writeup consisting of an exploitability-analysis report and, when available, a corresponding PoC. This information provides the LLM with the context of how the incorrect writeup was generated.
Second, the failure feedback includes two pieces of information: (1) the corresponding expert-written exploitability-analysis writeup and (2) an explicit indication that the LLM-generated writeup is incorrect. Together, they provide the LLM with the ground truth and a failure signal, supporting failure-driven self-reflection~\cite{xu2026beyond}.
Finally, the meta-prompt ($p_{{evi}^{meta}}$) defines the objective of the LLM’s inference task (shown in Table~\ref{table: p_evi_meta}). Rather than asking the LLM to identify the specific protocol evidence needed for the current target field, $p_{{evi}^{meta}}$ instructs the LLM to infer a general evidence-collection guidance.

\begin{table}[t]
\centering
\caption{Template of $p_{\mathrm{evi}}^{\mathrm{meta}}$.}
\label{table: p_evi_meta}
\vspace{-3pt}

\begingroup
\footnotesize
\normalfont
\setlength{\tabcolsep}{0pt}
\renewcommand{\arraystretch}{1}

\newcommand{\fieldplaceholder}[1]{{\normalfont\fontfamily{bch}\selectfont #1}}
\newcommand{\textindent}{0.3em}
\newcommand{\failureindent}{1.2em}
\newcommand{\historyindent}{1em}

\begin{tabular}{@{}>{\raggedright\arraybackslash}p{\columnwidth}@{}}
\toprule

{\normalfont\fontfamily{fvm}\selectfont
{\leftskip=\textindent

\noindent I asked an LLM to analyze the exploitability of some 3GPP Diameter field paths.
The LLM followed the exploitability-analysis prompt, $p_{\mathrm{exp}}$, to analyze the target field paths.
However, the LLM-generated exploitability analysis did not pass expert review.
I believe the failures come from the LLM not knowing enough protocol context about each target field.

\par}
}

\par\smallskip

{\normalfont\fontfamily{fvm}\selectfont
{\leftskip=\failureindent
\noindent\textbf{Failure case:}\par\vspace{0.2em}
\noindent\hspace*{\historyindent}%
\begin{tabular}{@{}l@{}}
Analysis History: \\
\{\texttt{target field and protocol evidence}\} \\
\{\texttt{exploitability-analysis prompt} $p_{\mathrm{exp}}$\} \\
\{\texttt{LLM-generated Writeup}\} \\
Failure Feedback: \\
\{\texttt{expert-written writeup}\}\\
\end{tabular}
\par}
}

\par\smallskip

{\normalfont\fontfamily{fvm}\selectfont
{\leftskip=\textindent
\noindent Your task is to infer a general protocol evidence-collection prompt from these feedbacks. 
Given any Diameter field path as input, this prompt should instruct an LLM to search all the relevant 3GPP specifications and collect needed protocol context related to this field path for exploitability analysis and PoC construction, 
so that the generated context is sufficient to support reports reproducing the ground-truth reports. 
The prompt must be reusable on unseen Diameter field paths. 
Output only the inferred prompt.
\par}
}

\\
\bottomrule
\end{tabular}

\endgroup
\vspace{-5pt}
\end{table}
More specifically, it asks the LLM to infer a prompt (i.e., $p_{evi}$) that can collect the protocol evidence required for exploitability analysis, such that, when the collected evidence is combined with the target Diameter field recorded in the analysis history and $p_{exp}$, the resulting LLM-generated writeup can reproduce the corresponding ground-truth writeup.

Once inferred, $p_{evi}$ is integrated into the first evidence-collection stage of the workflow shown in Fig.~\ref{fig: Workflow of LENS}. For any target Diameter field, $p_{evi}$ is used to collect the protocol evidence required for exploitability analysis in advance, which is then provided, together with the target field, to the subsequent exploitability-analysis stage guided by $p_{exp}$.

\vspace{-6pt}
\subsection{Iterative Co-Inference of the Two Prompts}
\label{subsec: Iterative Co-Inference of the Two Prompts}
\vspace{-6pt}

As can be seen from the designs in Sections~\ref{subsec: p_exp infer} and~\ref{subsec: p_evi infer}, the inference of $p_{exp}$ and $p_{evi}$ is inherently interdependent. On the one hand, inferring $p_{exp}$ benefits from protocol evidence collected by $p_{evi}$. On the other hand, inferring $p_{evi}$ relies on exploitability-analysis failures generated under $p_{exp}$. As a result, the two prompts cannot be inferred independently. To address this interdependence, our idea is to jointly refine the two prompts through an iterative process: \name\ alternates between exploitability-analysis discipline inference and evidence-collection guidance inference, allowing failures generated under $p_{exp}$ to support the inference of $p_{evi}$, and evidence collected by $p_{evi}$ to support the inference of $p_{exp}$.

As the starting point of the iterative refinement process, \name\ first infers an initial $p_{exp}$. Initially, only the target fields and their corresponding expert-written writeups are available. Although protocol evidence is unavailable, these examples nevertheless provide information from which an initial exploitability-analysis discipline can be inferred.
Next, starting from the initial $p_{exp}$, \name\ applies it to analyze the exploitable and non-exploitable examples used to infer it, and compares the generated writeups with the corresponding expert-written analyses. Whenever the generated writeups fail to reproduce the expert-written analyses, \name\ performs a refinement round. In each refinement round, the resulting failures are used to infer $p_{evi}$, which collects protocol evidence for updating $p_{exp}$. The updated $p_{exp}$ is then applied to the same examples again, producing a new round of exploitability analysis. This process is repeated until the generated writeups successfully reproduce the corresponding expert-written analyses.
Specifically, a writeup is considered successfully reproduced if (i) the generated exploitability conclusion is consistent with the expert-written one, (ii) the generated analysis is semantically equivalent to the expert-written analysis, as determined by expert review, and (iii) when the target field is exploitable, the generated PoC achieves the effect described in the corresponding writeup.

In practice, we find that the co-inference process converges quickly and typically requires only one to two rounds.
After refinement, the resulting $p_{exp}$ and $p_{evi}$ together form the final exploitability-analysis discipline and evidence-collection guidance, which are then incorporated into the \name\ workflow (Fig.~\ref{fig: Workflow of LENS}) to guide exploitability analysis and PoC construction for previously unseen Diameter fields.

\vspace{-5pt}
\subsection{Security Impact Assessment}
\vspace{-5pt}

When a Diameter field is determined to be exploitable by \name, the finding indicates a protocol-level security risk in Diameter. Building on these protocol-level findings, we further examine whether practical implementations are actually affected by them. To this end, we use the generated PoCs to perform security-impact assessment on implementations and investigate whether the predicted impacts indeed occur in practice.

Specifically, we perform the assessment in a controlled and isolated testing environment. Before the generated PoC can be applied, deployment-specific parameters, such as attacker identities, target network-element identities\cite{3GPP_TS_23003}, and subscriber identities (e.g., IMSI or IMPI), are instantiated in the attack messages. The instantiated attack messages are then sent to the target network element following the \name-generated attack procedure. Once the attack messages are sent, we manually monitor the state of the victim subscriber and the network responses returned during the attack, and assess whether the observed outcome is consistent with the impact predicted in the exploitability-analysis report.
For attacks targeting subscriber services, we examine whether the victim subscriber exhibits the predicted behavior, such as losing network connectivity or becoming unable to access the intended services. For attacks that trigger information disclosure from the core network, we examine whether the returned Diameter messages contain the sensitive information predicted by the analysis. Detailed assessment results are presented in Section~\ref{subsec: implementation-level findings}.

\vspace{-5pt}
\section{Evaluation}
\label{sec: Evaluation}
\vspace{-5pt}

\name is designed to infer exploitability-analysis discipline and evidence-collection guidance, and then use them to enable reliable LLM-based Diameter exploitability analysis. To evaluate whether \name achieves this goal, we organize our evaluation around four questions:
(1) Can \name successfully infer exploitability-analysis discipline and evidence-collection guidance?
(2) How effective is \name?
(3) What makes \name effective?
(4) What configurations are needed for \name?
We answer these questions in the following.

\vspace{-5pt}
\subsection{Understanding the Inferred Guidance and Discipline}
\label{subsec: Understanding the Inferred Guidance and Discipline}

\begin{table}[t]
\centering
\caption{Representative excerpts from $p_{\mathrm{evi}}$.}
\label{table:p_evi_excerpts}
\vspace{-3pt}

\begingroup
\footnotesize
\normalfont
\setlength{\tabcolsep}{0pt}
\renewcommand{\arraystretch}{1}
\newcommand{\textindent}{0.3em}

\begin{tabular}{@{}>{\raggedright\arraybackslash}p{\columnwidth}@{}}
\toprule

{\normalfont\fontfamily{fvm}\selectfont
{\leftskip=\textindent
\noindent \# Evidence Requirement\par

\smallskip
\noindent Your output must be a field-specific protocol evidence dossier that is
sufficient for a later LLM to:\par}

\begin{enumerate}[
    label=\arabic*.,
    leftmargin=2.2em,
    labelsep=0.5em,
    itemsep=0pt,
    topsep=0.3em,
    parsep=0pt,
    font=\normalfont\fontfamily{fvm}\selectfont
]
    \item identify the normal sender and receiver
    \item identify what the receiver does with the field
    \item trace the concrete causal chain from this field to later protocol behavior, if any
    \item explain how the receiving node is expected to process this field when the message is accepted
    \item state whether this field can alter persistent, semi-persistent, or only transient state in the receiver
\end{enumerate}


\par\vspace{0.6em}
{\leftskip=\textindent
\noindent \# State and Authority Analysis\par

\smallskip
\noindent Determine what kind of authority this AVP has in protocol terms:\par}

\begin{enumerate}[
    label=\arabic*.,
    leftmargin=2.2em,
    labelsep=0.5em,
    itemsep=0.15em,
    topsep=0.3em,
    parsep=0pt,
    font=\normalfont\fontfamily{fvm}\selectfont
]
    \item Does changing this field itself plausibly alter operational behavior?
    \item Name the exact downstream procedures or messages that consume or reflect the field
    \item If later legitimate signaling would overwrite the value, explain how and when
\end{enumerate}
}



\\
\bottomrule
\end{tabular}

\endgroup
\vspace{-7pt}
\end{table}
\vspace{-5pt}


The effectiveness of \name relies on two inferred prompts, $p_{evi}$ for protocol-evidence collection and $p_{exp}$ for exploitability analysis. Before evaluating \name, it is therefore important to understand whether \name successfully infers exploitability-analysis discipline and evidence-collection guidance. To answer this question, \magentaadd{we manually inspect a pair of final inferred prompts produced by \name}, focusing on two aspects: whether the inferred prompts generalize beyond the examples used for prompt inference and whether their contents indeed correspond to exploitability-analysis discipline and evidence-collection guidance.

\vspace{2pt}\noindent\textbf{Generality of the inferred prompts.}
Since both $p_{evi}$ and $p_{exp}$ are inferred from given examples, a natural concern is whether the inferred prompts merely capture patterns specific to those examples. We therefore examine the inferred prompts and find that neither $p_{evi}$ nor $p_{exp}$ contains any information tied to particular Diameter fields, messages, or interfaces appearing in the examples used for prompt inference. Instead, both prompts are expressed independently of any specific field. This suggests that the inferred prompts are general rather than example-specific.

\vspace{2pt}\noindent\textbf{The inferred evidence-collection guidance.}
We next examine the content of the inferred $p_{evi}$ to understand whether it indeed provides guidance on what protocol evidence should be collected and how such evidence can be obtained. Our analysis suggests that it does. Table~\ref{table:p_evi_excerpts} illustrates two representative excerpts from $p_{evi}$. The first excerpt defines a high-level scope of protocol information that should be collected by the LLM, including not only the semantics of the target field, but also sender and receiver roles, subscriber or network state updates, authority relationships, and downstream telecom effects. The second excerpt guides the LLM to analyze the protocol authority of the target field and its potential influence on subscriber or network state, such as whether it acts as an authoritative control point and what state entries may be modified if the field is manipulated. Beyond these examples, the prompt $p_{evi}$ further provides a rich set of specific strategies for evidence collection, such as identifying downstream protocol effects, tracing state changes caused by the target field, and locating the protocol information required for PoC construction.

\begin{table}[t]
\centering
\caption{Representative excerpts from $p_{\mathrm{exp}}$.}
\label{table:p_exp_excerpts}
\vspace{-3pt}

\begingroup
\footnotesize
\normalfont
\setlength{\tabcolsep}{0pt}
\renewcommand{\arraystretch}{1}
\newcommand{\textindent}{0.3em}

\begin{tabular}{@{}>{\raggedright\arraybackslash}p{\columnwidth}@{}}
\toprule

{\normalfont\fontfamily{fvm}\selectfont
{\leftskip=\textindent
\noindent \# Required Analysis Method\par

\smallskip
\noindent Base your reasoning on the supplied protocol evidence. Infer and explain:\par}

\begin{enumerate}[
    label=\arabic*.,
    leftmargin=2.2em,
    labelsep=0.5em,
    itemsep=0.15em,
    topsep=0.3em,
    parsep=0pt,
    font=\normalfont\fontfamily{fvm}\selectfont
]
    \item Direction and receiver

    \begin{itemize}[
        label=-,
        labelindent=0pt,
        leftmargin=*,
        labelsep=0.5em,
        itemsep=0pt,
        topsep=0pt,
        parsep=0pt,
        font=\normalfont\fontfamily{fvm}\selectfont
    ]
        \item Is the field in a request or answer? 
    \end{itemize}

    \item Field semantics and placement
    \begin{itemize}[
        label=-,
        labelindent=0pt,
        leftmargin=*,
        labelsep=0.5em,
        itemsep=0pt,
        topsep=0pt,
        parsep=0pt,
        font=\normalfont\fontfamily{fvm}\selectfont
    ]
        \item Exact grouped-AVP path and where the field sits.
    \end{itemize}

    \item Receiver processing
    \begin{itemize}[
        label=-,
        labelindent=0pt,
        leftmargin=*,
        labelsep=0.5em,
        itemsep=0pt,
        topsep=0pt,
        parsep=0pt,
        font=\normalfont\fontfamily{fvm}\selectfont
    ]
        \item Is it authoritative input, just metadata, or only optional context?
    \end{itemize}

      \item State impact
    \begin{itemize}[
        label=-,
        labelindent=0pt,
        leftmargin=*,
        labelsep=0.5em,
        itemsep=0pt,
        topsep=0pt,
        parsep=0pt,
        font=\normalfont\fontfamily{fvm}\selectfont
    ]
        \item What is the scope: per subscriber, per APN, per bearer, per serving-node context, temporary only, etc.?
    \end{itemize}

      \item Downstream/procedural consequences
    \begin{itemize}[
        label=-,
        labelindent=0pt,
        leftmargin=*,
        labelsep=0.5em,
        itemsep=0pt,
        topsep=0pt,
        parsep=0pt,
        font=\normalfont\fontfamily{fvm}\selectfont
    ]
        \item What later procedures, decisions, or messages depend on this field?
    \end{itemize}

    \item Field-specific security relevance
    \begin{itemize}[
        label=-,
        labelindent=0pt,
        leftmargin=*,
        labelsep=0.5em,
        itemsep=0pt,
        topsep=0pt,
        parsep=0pt,
        font=\normalfont\fontfamily{fvm}\selectfont
    ]
        \item Explain whether changing this field itself plausibly changes operation/security.
    \end{itemize}
        \item Overwrite/recovery
    \begin{itemize}[
        label=-,
        labelindent=0pt,
        leftmargin=*,
        labelsep=0.5em,
        itemsep=0pt,
        topsep=0pt,
        parsep=0pt,
        font=\normalfont\fontfamily{fvm}\selectfont
    ]
        \item Is recovery immediate, eventual, or unlikely?
    \end{itemize}

    \item PoC feasibility
    \begin{itemize}[
        label=-,
        labelindent=0pt,
        leftmargin=*,
        labelsep=0.5em,
        itemsep=0pt,
        topsep=0pt,
        parsep=0pt,
        font=\normalfont\fontfamily{fvm}\selectfont
    ]
          \item If exploitable, determine the minimum coherent message structure needed to exercise the field.
    \end{itemize}
\end{enumerate}

}
\\
\bottomrule
\end{tabular}

\endgroup
\vspace{-7pt}
\end{table}

\vspace{2pt}\noindent\textbf{The inferred exploitability-analysis discipline.}
Unlike $p_{evi}$, which guides the collection of protocol evidence, $p_{exp}$ is intended to guide how the collected evidence should be used for exploitability analysis. We therefore examine the content of the inferred $p_{exp}$ and find that it indeed encodes exploitability-analysis discipline. Table~\ref{table:p_exp_excerpts} illustrates a representative excerpt from $p_{exp}$. The excerpt does not directly ask the LLM to determine whether a field is Exploitable or Non-Exploitable. Instead, it requires the LLM to first infer and use a series of protocol facts before reaching a conclusion. These facts include message direction, AVP semantics, receiver processing, state and authority relationships, downstream consequences, and PoC feasibility. These considerations closely align with those typically examined by security experts during exploitability analysis, consistent with the intended role of $p_{exp}$ in our design. Beyond the excerpt shown in Table~\ref{table:p_exp_excerpts}, the full prompt further expands these considerations into detailed decision criteria and PoC-generation requirements; the complete prompt can be found in our artifact~\cite{lens}.

Overall, these observations suggest that \name successfully infers generalizable evidence-collection guidance and exploitability-analysis discipline.

\begin{table}[t]
\centering
\caption{End-to-End effectiveness of \name.}
\label{table: end-to-end effectiveness of LENS}
\vspace{-3pt}

{
\scriptsize
\setlength{\tabcolsep}{2.5pt}
\renewcommand{\arraystretch}{1}

{
\resizebox{\columnwidth}{!}{%
\begin{tabular}{lrrrr}
\toprule
\textbf{Model}
& \textbf{Recall (\%)}
& \textbf{Precision (\%)}
& \textbf{Accuracy (\%)}
& \textbf{Fields correct in $\geq$4/5 runs} \\
\midrule

GPT
& 96.00\textpm{}4.18
& 88.94\textpm{}2.12
& 92.00\textpm{}2.09
& 37/40 \\

GLM
& 94.00\textpm{}4.18
& 78.73\textpm{}6.28
& 84.00\textpm{}5.18
& 31/40 \\

Claude
& 98.00\textpm{}2.74
& 86.03\textpm{}3.31
& 91.00\textpm{}2.85
& 35/40 \\

Gemini
& 96.00\textpm{}4.18
& 84.50\textpm{}5.17
& 89.00\textpm{}3.79
& 35/40 \\

\bottomrule
\end{tabular}%
}
}
}
\vspace{-5pt}
\end{table}

\vspace{-6pt}
\subsection{End-to-End Effectiveness}
\label{subsec: End-to-End Effectiveness}
\vspace{-6pt}

\name\ is designed to perform reliable field-level Diameter exploitability analysis. We therefore evaluate whether it can achieve this goal in practice. To answer this question, we manually construct an evaluation dataset consisting of 40 Diameter fields covering both exploitable and non-exploitable cases across two representative Diameter interfaces considered in this work, S6a and Cx. For each interface, we randomly sample Diameter fields and manually analyze them until obtaining a balanced set consisting of 10 exploitable and 10 non-exploitable fields, resulting in 20 fields from S6a and 20 fields from Cx.
Constructing such an evaluation dataset requires expert analysis of the Diameter specifications, often involving information distributed across multiple documents. For each field, we manually analyze the relevant specifications, determine whether it is exploitable under our threat model, and produce the corresponding exploitability analysis report and protocol-level PoC when applicable. Here, none of the examples used during prompt inference are included in our evaluation dataset.


\magentaadd{
On this dataset, we evaluate \name\ using four underlying LLMs: GPT-5.4, GLM-5.2, Claude Opus 4.6, and Gemini 3.1 Pro.
For each LLM, we independently execute the complete \name\ pipeline five times. Each run starts from the same fixed seed examples but re-infers and refines both $p_{exp}$ and $p_{evi}$ from scratch, followed by evidence collection and exploitability analysis for all 40 evaluation fields.
}
A result produced by \name\ is considered correct only if (i) the exploitability conclusion is correct, (ii) the \name-generated analysis is semantically equivalent to the reference analysis, and (iii) for exploitable fields, the \name-generated attack procedure and attack messages are consistent with the reference PoC. This criterion is identical to that used in Section~\ref{subsec: Iterative Co-Inference of the Two Prompts} for determining whether an LLM-generated exploitability-analysis writeup successfully reproduces the corresponding expert-written writeup.

\vspace{2pt}\noindent\textbf{Effectiveness of \name.}
\magentaadd{
Table~\ref{table: end-to-end effectiveness of LENS} presents the effectiveness of \name, reporting recall, precision, and accuracy as mean $\pm$ standard deviation across five runs. Across the four LLMs, \name\ achieves mean recall of 94.00\% to 98.00\%, mean precision of 78.73\% to 88.94\%, and mean accuracy of 84.00\% to 92.00\%. In particular, the consistently high recall shows that \name\ correctly identifies nearly all exploitable fields and successfully produces the corresponding exploitability-analysis writeups. Table~\ref{table:appendix-end-to-end-s6a-cx} in the Appendix additionally breaks down the results by Diameter interface, reporting performance separately for S6a and Cx.
}

\magentaadd{
Beyond these aggregate metrics, we further examine whether the same fields are correctly analyzed across repeated runs. As shown in Table~\ref{table: end-to-end effectiveness of LENS}, GPT correctly analyzes 37/40 fields in at least four of five runs; Claude and Gemini each correctly analyze 35/40, while GLM correctly analyzes 31/40. These results show that, for every LLM, \name\ repeatedly produces correct analyses for most benchmark fields across independent runs. A more detailed breakdown of this run-to-run consistency is provided in Appendix Table~\ref{table:per-field-correctness}.
}

\magentaadd{
To understand the errors made by \name, we further manually inspect every field that is misclassified in any run of any LLM. Across 800 analyses (40 fields $\times$ 4 LLMs $\times$ 5 runs), \name\ makes 88 misclassifications across 16 fields: 72 false positives and 16 false negatives. We find that the 72 false positives fall into three recurring categories: attributing a message-level security effect to the target field, combining separate protocol behaviors into an invalid exploit chain, and overlooking protocol constraints because the collected evidence is incomplete. By contrast, all 16 false negatives share the same cause: \name\ assumes additional receiver-side authorization checks that are not required by the protocol specifications, causing feasible exploits to be rejected.
}

\begin{table}[t]
\centering
\caption{Runtime and cost of \name.}
\label{table: Efficiency and Cost of LENS}
\vspace{-3pt}

{
\footnotesize
\setlength{\tabcolsep}{3.5pt}
\renewcommand{\arraystretch}{1}

{
\begin{tabular*}{\columnwidth}
{@{\hspace{\tabcolsep}\extracolsep{\fill}}lrr@{\hspace{\tabcolsep}}}
\toprule

\textbf{Model}
& \multicolumn{1}{c}{\textbf{Runtime (min/field)}}
& \multicolumn{1}{c}{\textbf{Cost (USD/field)}} \\

\midrule

GPT
& 1.81\textpm{}0.31
& 0.230\textpm{}0.014 \\

GLM
& 2.15\textpm{}0.10
& 0.148\textpm{}0.002 \\

Claude
& 1.92\textpm{}0.12
& 0.298\textpm{}0.045 \\

Gemini
& 1.77\textpm{}0.10
& 0.234\textpm{}0.006 \\

\bottomrule
\end{tabular*}
}
}
\vspace{-5pt}
\end{table}
\vspace{2pt}\noindent\textbf{Efficiency and cost of \name.}
\magentaadd{
Since $p_{evi}$ and $p_{exp}$ are constructed once and then reused across target fields, we evaluate the runtime and monetary cost of analyzing each target Diameter field. Table~\ref{table: Efficiency and Cost of LENS} reports both as mean $\pm$ standard deviation over the same five runs used in the effectiveness evaluation for each LLM. Across the four LLMs, analyzing one field requires 1.77 to 2.15 minutes on average and incurs an average monetary cost of \$0.148 to \$0.298. These results indicate that \name\ can analyze Diameter fields with low per-field runtime and monetary cost across different LLMs.
}

\vspace{-6pt}
\subsection{Understanding \name Effectiveness}
\vspace{-6pt}
To understand why \name\ achieves high exploitability-analysis performance, we further evaluate the effectiveness of its key design choices. Using the same evaluation dataset and evaluation criterion as in Section~\ref{subsec: End-to-End Effectiveness}, we investigate two questions: (1) whether reliable exploitability analysis requires both exploitability-analysis discipline and protocol evidence, and (2) whether the proposed guidance-inference design is necessary and effective compared with simpler alternatives.
\magentaadd{These controlled studies use GPT-5.4 and the corresponding initial and refined prompts obtained from its first run in Section~\ref{subsec: End-to-End Effectiveness}.}

\vspace{2pt}\noindent\textbf{Effectiveness of discipline and evidence}.
To answer the first question, we evaluate three configurations of increasing capability: (1) Simple Prompt, which performs exploitability analysis without any guidance; (2) Initial $p_{exp}$, which performs exploitability analysis using the inferred exploitability-analysis discipline; and (3) Initial $p_{exp}$ $+$ Initial $p_{evi}$, which further provides protocol evidence collected using the inferred evidence-collection guidance. Here, we intentionally use the initial inferred prompts before refinement, as our goal is to understand the contributions of exploitability-analysis discipline and protocol evidence themselves.

\begin{table}[t]
\centering
\caption{Effectiveness of discipline and evidence in \name.}
\label{table: Effectiveness of Discipline and Evidence}
\vspace{-3pt}
\setlength{\tabcolsep}{1.1pt}
\renewcommand{\arraystretch}{1}


{\footnotesize\itshape Discipline: exploitability-analysis discipline; Evidence: protocol evidence.\par}
\vspace{2pt}

\begin{adjustbox}{max width=\columnwidth,center}
{\fontsize{9.3pt}{10.1pt}\selectfont
\begin{tabular}{@{\hspace{2.5pt}}lccrrr@{\hspace{2.5pt}}}
\toprule
\textbf{Configuration}
& \textbf{Discipline}
& \textbf{Evidence}
& \multicolumn{1}{c}{\textbf{Recall}}
& \multicolumn{1}{c}{\textbf{Precision}}
& \multicolumn{1}{c}{\textbf{Accuracy}} \\
\midrule
Simple Prompt
& \xmark & \xmark & 30.00\% & 37.50\% & 40.00\% \\
Initial $p_{exp}$
& \cmark & \xmark & 40.00\% & 80.00\% & 65.00\% \\
Initial $p_{exp}$ + Initial $p_{evi}$
& \cmark & \cmark & 80.00\% & 84.21\% & 82.50\% \\
\bottomrule
\end{tabular}
}
\end{adjustbox}
\vspace{-7pt}
\end{table}

Table~\ref{table: Effectiveness of Discipline and Evidence} presents the results. The Simple Prompt configuration achieves only 30.00\% recall, 37.50\% precision, and 40.00\% accuracy. Providing exploitability-analysis discipline (i.e., Initial $p_{exp}$) substantially improves performance, increasing accuracy from 40.00\% to 65.00\% and more than doubling precision from 37.50\% to 80.00\%, while recall improves modestly. These results suggest that exploitability-analysis discipline plays an important role in reliable LLM-based Diameter exploitability analysis, particularly through improving precision.
Adding protocol evidence (i.e., Initial $p_{evi}$) further improves performance. Compared with using exploitability-analysis discipline alone, providing protocol evidence increases accuracy from 65.00\% to 82.50\% and doubles recall from 40.00\% to 80.00\%, while still achieving high precision. These results suggest that protocol evidence also plays an important role in reliable LLM-based Diameter exploitability analysis, particularly in improving recall.

Overall, the results suggest that both exploitability-analysis discipline and protocol evidence are necessary for reliable LLM-based Diameter exploitability analysis. They play complementary roles, with the former primarily improving precision and the latter primarily improving recall.

\vspace{2pt}\noindent\textbf{Effectiveness of guidance-inference design}.
Having established the importance of exploitability-analysis discipline and protocol evidence, we next evaluate whether the corresponding design choices in \name\ are more effective than simpler alternatives. Specifically, we evaluate three components of our design: exploitability-analysis discipline inference (Section~\ref{subsec: p_exp infer}), evidence-collection guidance inference (Section~\ref{subsec: p_evi infer}), and guidance refinement (Section~\ref{subsec: Iterative Co-Inference of the Two Prompts}).

\vspace{1pt}\noindent~$\bullet$~\textbf{\textit{Exploitability-analysis discipline inference}}.
As discussed in Section~\ref{subsec: p_exp infer}, existing exploit writeups reflect the tacit exploitability-analysis discipline used by security experts. Therefore, to make such discipline available to LLMs for reliable exploitability analysis, \name\ infers the discipline from exploit writeups using the procedure illustrated in Figure~\ref{fig: p_exp}. An immediate question is whether such inference is necessary, or whether directly providing the same exploit writeups as few-shot demonstrations would be sufficient. To evaluate the effectiveness of discipline inference, we compare it against a few-shot baseline that uses the same exploitable and non-exploitable examples used for discipline inference.
Table~\ref{table: Effectiveness of the Guidance-Inference Design} presents the results. The few-shot baseline achieves 77.78\% precision and 62.50\% accuracy, suggesting that directly providing exploit writeups allows the LLM to partially leverage the exploitability-analysis discipline embedded in them. However, the inferred exploitability-analysis discipline achieves higher precision and accuracy, improving them to 100.00\% and 72.50\%, respectively, while providing a modest improvement in recall. These results suggest that inferring exploitability-analysis discipline is more effective than directly using the same examples as demonstrations, supporting the design choice of discipline inference in \name.

\vspace{1pt}\noindent~$\bullet$~\textbf{\textit{Evidence-collection guidance inference}}. 
As discussed in Section~\ref{subsec: p_evi infer}, the protocol evidence required for exploitability analysis cannot be identified in advance. To address this challenge, \name\ learns evidence-collection guidance from exploitability-analysis failures using the procedure illustrated in Figure~\ref{fig: p_evi}. A natural question is whether such guidance inference is necessary, or whether the required protocol evidence can be collected using simpler retrieval-based approaches. To evaluate the effectiveness of evidence-collection guidance inference, we compare it against three retrieval-based baselines following the standard RAG paradigm~\cite{gao2023retrieval}: (1) Keyword Retrieval, which collects all specification sections containing the target field through keyword matching; (2) Field-Relevant Retrieval, which prompts the LLM to collect information generally related to the target field; and (3) Exploitability-Relevant Retrieval, which prompts the LLM to collect information specifically relevant to exploitability analysis of the target field.

\begin{table}[t]
\centering
\caption{Effectiveness of the guidance-inference design.}
\label{table: Effectiveness of the Guidance-Inference Design}
\vspace{-3pt}
\fontsize{7.5pt}{8.6pt}\selectfont
\setlength{\tabcolsep}{0pt}
\renewcommand{\arraystretch}{1}

\begin{tabularx}{\columnwidth}{
@{}
>{\raggedright\arraybackslash}p{0.18\columnwidth}
@{\hspace{7pt}}
>{\raggedright\arraybackslash}X
@{\hspace{5.5pt}}
r
@{\hspace{4pt}}
r
@{\hspace{4pt}}
r
@{}
}
\toprule
\textbf{Component} & \textbf{Configuration}
& \textbf{Recall}
& \textbf{Precision}
& \textbf{Accuracy} \\
\midrule
\multirow{2}{*}{\makecell[l]{$p_{exp}$\\Inference}}
& Few-shot Examples                         & 35.00\%  & 77.78\%  & 62.50\% \\
& Inferred $p_{exp}$                        & 45.00\%  & 100.00\% & 72.50\% \\
\midrule
\multirow{4}{*}{\makecell[l]{$p_{evi}$\\Inference}}
& Keyword Retrieval                         & 55.00\%  & 73.33\%  & 67.50\% \\
& Field-Relevant Retrieval                  & 40.00\%  & 88.89\%  & 67.50\% \\
& \begin{adjustbox}{max width=\linewidth}
Exploitability-Relevant Retrieval
\end{adjustbox}
                                                & 50.00\%  & 83.33\%  & 70.00\% \\
& Inferred $p_{evi}$                        & 80.00\%  & 84.21\%  & 82.50\% \\
\midrule
\multirow{2}{*}{\makecell[l]{Joint\\Refinement}}
& Initial $p_{exp}$ + Initial $p_{evi}$     & 80.00\%  & 84.21\%  & 82.50\% \\
& \begin{adjustbox}{max width=\linewidth}
Refined $p_{exp}$ + Refined $p_{evi}$
\end{adjustbox}
                                                & 100.00\% & 90.91\%  & 95.00\% \\
\bottomrule
\end{tabularx}
\vspace{-7pt}
\end{table}
Table~\ref{table: Effectiveness of the Guidance-Inference Design} presents the results. Using the protocol evidence collected by the retrieval-based baselines, exploitability analysis achieves recall ranging from 40.00\% to 55.00\% and accuracy ranging from 67.50\% to 70.00\%. In contrast, using the protocol evidence collected by the inferred $p_{evi}$, exploitability analysis achieves 80.00\% recall and 82.50\% accuracy, substantially outperforming all retrieval-based baselines while maintaining comparable precision. These results suggest that how protocol evidence is collected has a substantial impact on exploitability-analysis performance. In particular, learning evidence-collection guidance from exploitability-analysis failures is more effective than retrieval-based evidence collection, supporting the design choice of evidence-collection guidance inference in \name.

\vspace{1pt}\noindent~$\bullet$~\textbf{\textit{Refinement of $p_{exp}$ and $p_{evi}$}}. 
As discussed in Section~\ref{subsec: Iterative Co-Inference of the Two Prompts}, the inference of $p_{exp}$ and $p_{evi}$ depends on each other, as exploitability-analysis discipline affects evidence collection, while collected evidence in turn affects discipline inference. Therefore, the initially inferred prompts are further refined through iterative refinement. To evaluate the effectiveness of this design, we compare the initially inferred prompts against the final refined prompts obtained after refinement.
Table~\ref{table: Effectiveness of the Guidance-Inference Design} presents the results. Compared with initial $p_{exp}$ $+$ initial $p_{evi}$, refinement substantially improves performance across all evaluation metrics. Most notably, refinement increases recall from 80.00\% to 100.00\%, meaning that all exploitable fields in the evaluation dataset are successfully identified. Meanwhile, precision and accuracy are further improved, reaching 90.91\% (from 84.21\%) and 95.00\% (from 82.50\%), respectively. These results demonstrate that iterative refinement further enhances the effectiveness of the inferred guidance, enabling highly reliable LLM-based Diameter exploitability analysis.

\vspace{-6pt}
\subsection{Configuration Study}
\label{subsec: configuration study}
\vspace{-6pt}

Having established the effectiveness of \name and its key design choices, we next study the configurations of \name. In particular, \name involves several configurable choices, including the number of examples used for prompt inference, the number of co-inference refinement rounds, and the information used for evidence-collection guidance inference. Studying these choices helps identify the configurations needed for reliable exploitability analysis. We therefore investigate two questions: (1) how many examples are needed for prompt inference, and (2) what information is needed for evidence-collection guidance inference.
\magentaadd{We use GPT-5.4 throughout this section.}

\begin{figure}[t]
  \centering

  \includegraphics[width=1\linewidth]{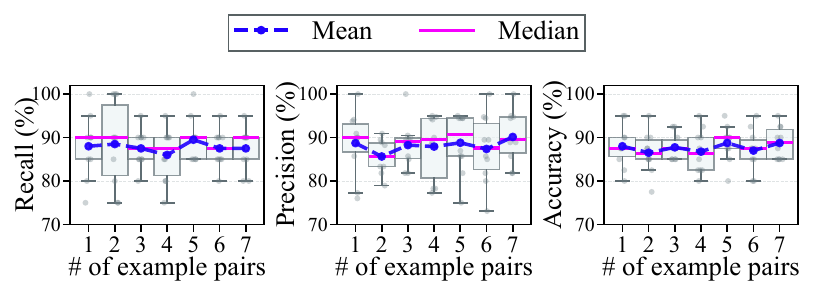}
  \caption{Effect of the number of example pairs on prompt inference.}
  \label{fig: Effect of the Number of Examples on Prompt Inference}
  \vspace{-6pt}
\end{figure}

\vspace{2pt}\noindent\textbf{Number of examples}.
As discussed in Section~\ref{subsec: p_exp infer}, \name infers its exploitability-analysis prompt from examples constructed from expert-written exploit writeups. We first study how many such examples are needed for prompt inference. To answer this question, we construct a prompt-inference dataset consisting of eight previously reported exploitable Diameter cases~\cite{kotte2016analysis,tan2021device,mashukov2017diameter, holtmanns2018mobile} and eight manually constructed non-exploitable cases. This dataset is used only for prompt inference and is separate from the evaluation dataset used in Section~\ref{subsec: End-to-End Effectiveness}. On this dataset, we evaluate how the number of examples used for prompt inference affects the resulting exploitability-analysis performance. Specifically, we start with one exploitable case and one non-exploitable case, and gradually increase the number to seven exploitable cases and seven non-exploitable cases. For each setting, we randomly select ten different combinations of exploitable and non-exploitable cases from the prompt-inference dataset, infer the corresponding prompts, and evaluate them on the evaluation dataset used in Section~\ref{subsec: End-to-End Effectiveness}. Figure~\ref{fig: Effect of the Number of Examples on Prompt Inference} presents the results.

Across all settings, the inferred prompts achieve comparable recall, precision, and accuracy. Although using more examples occasionally leads to slight improvements in these metrics, the gains are marginal. Overall, we do not observe a consistent trend that increasing the number of examples improves exploitability-analysis performance.
Taken together, these observations suggest that reliable prompt inference can be achieved with only a small number of examples. In particular, we do not observe clear benefits from using more than one exploitable case and one non-exploitable case. \name therefore adopts this configuration throughout the paper.

\begin{table}[t]
\centering
\caption{Effect of information used for $p_{evi}$ inference.}
\label{table: Effect of information}
\vspace{-3pt}
\setlength{\tabcolsep}{2pt}
\resizebox{\columnwidth}{!}{%
\begin{tabular}{@{\hspace{3.5pt}}cccrrr@{\hspace{3.5pt}}}
\toprule
\makecell{\textbf{LLM-Generated}\\\textbf{Writeup}} &
\makecell{\textbf{Expert-Written}\\\textbf{Writeup}} &
\makecell{\textbf{Analysis}\\\textbf{History}} &
\textbf{Recall} &
\textbf{Precision} &
\textbf{Accuracy} \\
\midrule
\cmark & \xmark & \xmark & 55.00\% & 84.62\% & 72.50\% \\
\cmark & \cmark & \xmark & 70.00\% & 87.50\% & 80.00\% \\
\cmark & \cmark & \cmark & 80.00\% & 84.21\% & 82.50\% \\
\bottomrule
\end{tabular}%
}
\vspace{-7pt}
\end{table}

\vspace{2pt}\noindent\textbf{Information for $p_{evi}$ inference}.
As discussed in Section~\ref{subsec: p_evi infer}, \name infers $p_{evi}$ using three types of information obtained from a failed exploitability analysis: (1) the LLM-generated writeup together with a failure signal indicating that the analysis did not pass expert review, (2) the corresponding expert-written writeup as ground truth, and (3) the analysis history describing how the incorrect writeup was produced. Next, we study whether each type of information contributes to $p_{evi}$ inference. To this end, we fix the initial $p_{exp}$ \magentaadd{obtained from the first GPT-5.4 run in Section~\ref{subsec: End-to-End Effectiveness}} and progressively provide these information sources for $p_{evi}$ inference. For each setting, we infer the resulting $p_{evi}$ and evaluate it on the evaluation dataset used in Section~\ref{subsec: End-to-End Effectiveness}. 
As shown in Table~\ref{table: Effect of information}, providing additional information consistently improves performance. Starting from the LLM-generated writeup and failure signal alone, adding the expert-written writeup substantially improves all three metrics, increasing recall from 55.0\% to 70.0\%, precision from 84.6\% to 87.5\%, and accuracy from 72.5\% to 80.0\%. Further providing the analysis history continues to improve performance, increasing recall from 70.0\% to 80.0\% and accuracy from 80.0\% to 82.5\%, while causing only a small decrease in precision. Taken together, these results justify the use of all three types of information for $p_{evi}$ inference.

\vspace{-5pt}
\section{Findings}
\vspace{-5pt}

Section~\ref{sec: Evaluation} demonstrates the effectiveness of \name\ for Diameter exploitability analysis. We now apply \name\ to the Diameter protocol at scale to understand its security landscape at both the protocol and implementation levels. \magentaadd{Using GPT-5.4 with the  $p_{evi}$ and $p_{exp}$ from the first run in Section~\ref{subsec: End-to-End Effectiveness},} we first investigate which Diameter fields are exploitable at the protocol level under our threat model, and then examine how prevalent these protocol-level weaknesses are in practical implementations. The following subsections present protocol-level findings, implementation-level findings, and representative attacks revealed by our analysis.

\begin{table}[t]
\centering
\caption{Protocol-level exploitability findings.}
\label{table: Protocol-level exploitability findings}
\vspace{-3pt}

{
\footnotesize
\setlength{\tabcolsep}{3.5pt}
\renewcommand{\arraystretch}{1}

{
\begin{tabular*}{\columnwidth}
{@{\hspace{\tabcolsep}\extracolsep{\fill}}lrrrr@{\hspace{\tabcolsep}}}
\toprule

\textbf{Interface}
& \textbf{Total Fields}
& \textbf{LENS-identified}
& \textbf{Validated}
& \textbf{New} \\

\midrule

S6a
& 600
& 176 (29.33\%)
& 160 (26.67\%)
& 152 \\

Cx
& 166
& 47 (28.31\%)
& 44 (26.51\%)
& 44 \\

\textbf{Overall}
& 766
& 223 (29.11\%)
& 204 (26.63\%)
& 196 \\

\bottomrule
\end{tabular*}
}
}
\vspace{-7pt}
\end{table}

\vspace{-6pt}
\subsection{Protocol-Level Findings}
\label{subsec: protocol-level findings}
\vspace{-6pt}

To understand the security landscape of Diameter, we use \name\ to analyze the S6a and Cx interfaces, which are responsible for LTE subscriber management and authentication, and IMS service control (e.g., voice-call services), respectively~\cite{29272, 29229}. This analysis covers 766 fields, including 600 fields in S6a and 166 fields in Cx.

\magentaadd{
As shown in Table~\ref{table: Protocol-level exploitability findings}, \name\ initially identifies 223 fields as exploitable. To verify whether these fields are indeed exploitable at the protocol level, we manually audit all 223 fields. For each field, we check whether the collected evidence is sufficient and verifiable and whether it supports the claimed field-specific exploitability. This audit confirms 204 fields. The other 19 do not pass the audit for reasons consistent with the failure modes observed in Section~\ref{subsec: End-to-End Effectiveness}. Accordingly, using \name, we identify 204 validated protocol-level exploitability findings. This means that over one quarter of the 766 analyzed Diameter fields (26.63\%) are exploitable under our threat model, revealing widespread field-level exploitability once Diameter network elements become reachable.
}
More importantly, \magentaadd{196} of these exploitable fields (\magentaadd{96.08\%}) have not been reported in prior academic literature or public attack reports.

To further understand the security consequences associated with these exploitable fields, we analyze their corresponding protocol-level PoCs. In total, \name generates \magentaadd{222} protocol-level PoCs for the \magentaadd{204 validated exploitable fields}. The number of generated PoCs exceeds the number of exploitable fields because the same field may lead to different behaviors when assigned different values, resulting in multiple distinct security consequences.
After analyzing the generated PoCs, we find that they lead to diverse security consequences. Table~\ref{table: security consequences} summarizes our categorization of these consequences. The results reveal a clear concentration around service availability: the most common consequences are service disruption and service degradation, accounting for \magentaadd{85 (38.29\%) and 76 (34.23\%)} PoCs, respectively. Service disruption can completely disable subscriber services, such as voice calls, SMS, or Internet access, while service degradation can significantly reduce service quality, for example by forcing subscriber network traffic to extremely low data rates (e.g., 1 bit/s). These two categories account for nearly three quarters (\magentaadd{72.52\%}) of all generated PoCs, indicating that most identified threats ultimately affect the availability of subscriber services.
Beyond impacts on service availability, we also observe additional security and privacy consequences that directly affect subscribers. 
In particular, 32 PoCs (\magentaadd{14.41\%}) trigger information leakage, exposing sensitive subscriber information such as authentication vectors and subscriber-profile data (e.g., phone numbers and service configurations).
Another 15 PoCs (\magentaadd{6.76\%}) cause accelerated battery depletion by forcing victim devices to repeatedly perform network procedures and consume excessive power.
In addition, 14 PoCs (\magentaadd{6.31\%}) enable precise cell-level subscriber tracking or monitoring of subscriber online status.
These findings show that, although the exploitable fields reside in core-network signaling protocols, their impacts ultimately affect subscribers, threatening not only service availability, but also subscriber privacy and device usability.


\vspace{2pt}\noindent
\magentaadd{\textbf{Field-level and message-level views.}}
\magentaadd{Since \name\ analyzes exploitability at the granularity of individual Diameter fields, the findings above are reported at the field level. To provide a complementary view of these findings, we further characterize the 204 validated fields by the Diameter message, receiver, and receiver-side operation involved. In total, we identify 49 distinct scenarios. For example, 18 exploitable fields associated with the IDR message can each be manipulated by an attacker impersonating the HSS to update the UE's APN/PDN connectivity configuration at the MME, leading to service disruption or service degradation. Appendix Table~\ref{table: message-level findings} presents the complete results.}

\begin{table}[t]
\centering
\caption{Security consequences of protocol-level PoCs.}
\label{table: security consequences}
\vspace{-3pt}

{
\footnotesize
\setlength{\tabcolsep}{4.5pt}
\renewcommand{\arraystretch}{1}

{
\begin{tabular*}{\columnwidth}{
@{\hspace{\tabcolsep}}
c
@{\hspace{1.7em}}
l
@{\extracolsep{\fill}}
r
@{\hspace{1.7em}}
r
@{\hspace{1.7em}}
}
\toprule

\textbf{CID}
& \textbf{Consequence Category}
& \textbf{PoCs}
& \textbf{Percentage} \\

\midrule

C1
& Service Disruption
& 85
& 38.29\% \\

C2
& Service Degradation
& 76
& 34.23\% \\

C3
& Information Leakage
& 32
& 14.41\% \\

C4
& Battery Depletion
& 15
& 6.76\% \\

C5
& Subscriber Tracking
& 14
& 6.31\% \\

\midrule

\multicolumn{2}{l}{\textbf{Total}}
& \textbf{222}
& \textbf{100.00\%} \\

\bottomrule
\end{tabular*}
}
}
\vspace{-7pt}
\end{table}

\vspace{-6pt}
\subsection{Implementation-Level Findings}
\label{subsec: implementation-level findings}
\vspace{-6pt}

\magentaadd{To examine whether the protocol-level findings identified above also hold in practical implementations,} we conduct implementation-level assessment in four testing environments, including two open-source core-network implementations (Open5GS~\cite{open5gs2026} and Kamailio~\cite{kamailio2026}) and two anonymized commercial core-network implementations, denoted as Commercial \textit{A} and Commercial \textit{B}. Open5GS supports the S6a interface, Kamailio supports the Cx interface, Commercial \textit{A} supports both S6a and Cx, and Commercial \textit{B} supports Cx. All assessments are conducted in isolated environments separated from operational networks.

\begin{table}[t]
\centering
\caption{Implementation-level assessment results.}
\label{table: implementation-level impacts} 
\vspace{-3pt}

\footnotesize
\setlength{\tabcolsep}{2.5pt}
\renewcommand{\arraystretch}{1}

{\footnotesize\itshape
Impl.: implementation; Intf.: interface; EF: exploitable fields;\\
T: type (O: open-source; C: commercial).\par}
\vspace{2pt}

\resizebox{\columnwidth}{!}{%
\begin{tabular}{@{}l@{\hspace{8pt}}cc@{\hspace{5pt}}c@{\hspace{1pt}}c@{\hspace{1pt}}c@{\hspace{1pt}}c@{}}
\toprule
\textbf{Impl.} &
\textbf{T} &
\textbf{Intf.} &
\makecell[c]{\textbf{Supported}\\\textbf{EF}} &
\makecell[c]{\textbf{Confirmed}\\\textbf{EF(\%)}} &
\makecell[c]{\textbf{Associated}\\\textbf{PoCs}} &
\makecell[c]{\textbf{Confirmed}\\\textbf{PoCs(\%)}} \\
\midrule
Open5GS
& O & S6a
& 25 & \phantom{0}4 \phantom{0}(16.00\%) & 28 & \phantom{0}7 (25.00\%) \\

Kamailio
& O & Cx
& \phantom{0}9 & \phantom{0}9 (100.00\%) & 15 & 10 (66.67\%) \\

\addlinespace[1pt]
Commercial \textit{A}
& C & S6a
& 40 & 35 \phantom{0}(87.50\%) & 46 & 40 (86.96\%) \\

Commercial \textit{A}
& C & Cx
& 16 & 16 (100.00\%) & 24 & 23 (95.83\%) \\

Commercial \textit{B}
& C & Cx
& 16 & 15 \phantom{0}(93.75\%) & 18 & 15 (83.33\%) \\

\midrule
\multicolumn{3}{@{}l}{\textbf{Overall}}
& \textbf{56}
& \textbf{51 \phantom{0}(91.07\%)}
& \textbf{71}
& \textbf{64 (90.14\%)} \\
\bottomrule
\end{tabular}%
}
\vspace{-7pt}
\end{table}

Table~\ref{table: implementation-level impacts} summarizes the results. Among the \magentaadd{204} protocol-level exploitable fields identified in Section~\ref{subsec: protocol-level findings}, only 56 are supported by at least one evaluated implementation. The difference arises because many Diameter fields are optional, and different implementations support different subsets of protocol functionalities.
Among these 56 supported fields, we find that 51 (91.07\%) are confirmed to be exploitable in at least one implementation, while 64 (90.14\%) of their corresponding 71 protocol-level PoCs successfully trigger the predicted security impacts. 
The high field-confirmation rate (91.07\%) and PoC-success rate (90.14\%) indicate that practical implementations are highly likely to be affected if the corresponding protocol functionality is implemented.

To understand why some supported protocol-level weaknesses do not produce the predicted impacts, we further analyze the unsuccessful cases. Since the source code of the commercial implementations is unavailable, we focus our analysis on the open-source implementations. Our analysis reveals that the unsuccessful cases stem from two implementation-specific decisions.
In Open5GS, all unsuccessful cases are caused by the same implementation-specific behavior. When a UE re-attaches to the network, Open5GS always retrieves fresh subscriber data from the HSS and overwrites the subscriber state previously manipulated by the PoC in the MME, thereby eliminating the attack effects. This behavior is not required by the 3GPP specifications\cite{29272}, but is instead an implementation-specific design choice made by Open5GS.
In Kamailio, all unsuccessful cases arise because the implementation supports the target fields but does not implement the functionalities associated with the manipulated field values. As a result, setting these fields to the manipulated values cannot trigger the expected protocol behaviors or security impacts.

We further analyze the concrete impacts caused by the 64 PoCs observed in our evaluated implementations. We find that these impacts cover all security-consequence categories identified in Section~\ref{subsec: protocol-level findings}. For example, the observed impacts include complete LTE service loss, exposure of subscriber authentication vectors, and precise cell-level subscriber tracking. The detailed impacts are summarized in Table~\ref{table: 5.2_impact}, while Appendix Table~\ref{table: affected-diameter-fields} provides a complete list of the corresponding fields, interfaces, impact categories, and affected implementations for all 64 confirmed PoCs.

\begin{table}[t]
\centering
\caption{Security impacts of confirmed PoCs.}
\label{table: 5.2_impact}
\renewcommand{\arraystretch}{1}
\vspace{-3pt}

\footnotesize
\setlength{\tabcolsep}{1.2pt}

\newlength{\rightinset}
\setlength{\rightinset}{0.25em}

\newlength{\vulncolwd}
\settowidth{\vulncolwd}{\textbf{PoCs}}
\addtolength{\vulncolwd}{0.2em}

\newlength{\numwd}
\settowidth{\numwd}{00}

\newlength{\totallabelwd}
\settowidth{\totallabelwd}{\textit{Total}\hspace{0.3em}}

\newcommand{\vnum}[1]{%
  \makebox[\linewidth][r]{%
    \makebox[\numwd][r]{#1}%
  }%
}

\newcommand{\totalnum}[1]{%
  \makebox[\linewidth][r]{%
    \makebox[\totallabelwd][l]{\textit{Total}\hspace{0.3em}}%
    \makebox[\numwd][r]{#1}%
  }%
}

\newcommand{\cid}[2]{%
  \multirow[t]{#1}{0.08\columnwidth}{\centering\arraybackslash #2}%
}

\begin{tabularx}{\columnwidth}{
@{}>{\centering\arraybackslash}p{0.08\columnwidth}
@{\hspace{1.4em}}
>{\raggedright\arraybackslash}X
@{\hspace{-0.6em}}
>{\raggedleft\arraybackslash}p{\vulncolwd}
@{\hspace{\rightinset}}}
\toprule
\textbf{CID} 
& \textbf{Impact}
& \makebox[\linewidth][r]{\textbf{PoCs}} \\
\midrule

\cid{3}{C1}
& Complete LTE service loss                                & \vnum{25} \\
& All voice and SMS services unavailable                   & \vnum{8} \\
& Incoming voice calls and SMS unavailable                 & \vnum{9} \\
& & \totalnum{42} \\

\midrule
\cid{2}{C2}
& Data rate reduced to extremely low levels                & \vnum{6} \\
& Only IPv6 services remain accessible                     & \vnum{2} \\
& & \totalnum{8} \\

\midrule
\cid{4}{C3}
& Exposure of subscriber profile information               & \vnum{2} \\
& Exposure of subscriber authentication vectors            & \vnum{2} \\
& Exposure of network-element capabilities                 & \vnum{4} \\
& Exposure of internal network configuration               & \vnum{2} \\
& & \totalnum{10} \\

\midrule
\cid{1}{C4}
& Accelerated battery drain                                & \vnum{1} \\
& & \totalnum{1} \\

\midrule
\cid{2}{C5}
& Precise cell-level location tracking                     & \vnum{1} \\
& Monitoring subscriber online status                      & \vnum{2} \\
& & \totalnum{3} \\

\bottomrule
\end{tabularx}
\vspace{-7pt}
\end{table}

Overall, we identify 95 previously unknown implementation-confirmed vulnerabilities, including 17 in the two open-source implementations and 78 in the two commercial implementations. This number exceeds the 64 PoCs whose predicted impacts are successfully observed in our evaluated implementations, as some PoCs affect multiple implementations.
We responsibly disclosed all identified vulnerabilities to the affected vendors and maintainers. Notably, both commercial vendors confirmed all reported vulnerabilities. At the time of writing, 5 CVE identifiers and 7 CNVD identifiers have been assigned to the reported Open5GS vulnerabilities, while the remaining cases are under review.

\vspace{-6pt}
\subsection{Representative Attacks}
\vspace{-6pt}

The protocol-level PoCs identified by \name already reveal powerful attack capabilities against cellular core networks. To further illustrate their practical implications, we present three representative attacks constructed in our study, all of which are previously unreported to the best of our knowledge and have been successfully demonstrated in our testbed.

\vspace{2pt}\noindent\textbf{Persistent rogue-network attachment}.
This attack enables an attacker to transparently take over a victim UE’s cellular connectivity, including data, voice, and messaging services, by constructing a rogue EPS/IMS network. Normally, such an attack would be prevented because a rogue network cannot complete UE-network mutual authentication without valid authentication material. However, by combining multiple information-leakage PoCs (PID-4, PID-5, PID-51, and PID-54 in Appendix Table~\ref{table: affected-diameter-fields}), the attacker can obtain both the required authentication material and subscriber profiles from the HSS, enabling the construction of a rogue EPS/IMS network. After constructing the rogue network, the attacker can leverage the service-disruption PoC (PID-2) to send a crafted \textit{Cancel-Location-Request (CLR)}, forcing the victim UE to detach from the legitimate network. When the UE subsequently performs network reselection, the attacker can attract it to the attacker-controlled network through a rogue eNB broadcasting a stronger signal. The attacker-controlled network can then use the stolen EPS-AKA and IMS-AKA authentication vectors to complete EPS attachment and IMS registration, and use the obtained subscriber profiles to provide cellular and IMS services that are indistinguishable from those of the legitimate network.

This attack highlights the severity of the information exposed by the identified Diameter vulnerabilities. The exposed authentication vectors and subscriber profiles are not merely pieces of subscriber information; together, they provide all critical materials required to construct a rogue EPS/IMS network. As a result, an attacker can impersonate a legitimate operator and gain full control over the victim's cellular connectivity, including data, voice, and messaging services.

\noindent\textbf{Persistent LTE service loss}.
This attack enables an attacker to cause persistent LTE service loss for a victim UE by modifying APN-related subscriber configuration stored in the MME. These configuration parameters are required for the MME to provide cellular data service to the UE, and invalid values can prevent the service from being successfully established. Using Diameter vulnerabilities that can modify APN-related configuration (e.g., PID-11 to PID-22), an attacker can replace legitimate values with invalid ones. Additionally, to immediately trigger the attack, the attacker can further leverage a service-disruption PoC (e.g., PID-2) to force the victim UE to perform a new attach procedure. During the subsequent attach procedure, the MME will use the modified configuration. Because the modified configuration is invalid, the MME fails to establish service for the UE, causing the victim UE to lose LTE service~\cite{3GPP_TS_23401}.

The impact of this attack can be particularly severe. Once the attack succeeds, the victim UE may remain without LTE service throughout the service area of the affected MME. Repeated device reboots or enabling and disabling airplane mode do not restore connectivity, because the serving MME continues to use the modified subscriber configuration. In our testbed, the victim remained disconnected until it moved outside the service area of the affected MME and attached through another MME.

\vspace{2pt}\noindent\textbf{Accelerated battery depletion.}
This attack enables an attacker to significantly accelerate battery depletion of a victim UE by modifying the periodic TAU timer delivered by the MME\cite{3GPP_TS_24301_T3412}. The timer determines how frequently the UE performs periodic TAU procedures to maintain its network registration. Using the timer-modification vulnerability (PID-41), an attacker can set the timer to only a few seconds. To trigger the MME to deliver the modified timer to the UE, the attacker can further leverage a service-disruption PoC (e.g., PID-2) to force the victim UE to perform a new attach procedure. This is because the modified timer is only delivered during network attachment procedures. As a result, after re-attaching to the network, the UE will repeatedly perform network signaling far more frequently than intended, significantly increasing battery consumption.

The impact of this attack can be particularly severe for battery-powered IoT devices, which often rely on long idle intervals to achieve extended battery lifetime. In our testbed, we set the timer to 2 seconds and evaluated the attack on two battery-powered IoT devices: a low-power edge gateway and a low-power data collector. The impact was substantial. For the former device, which is advertised with a three-year battery lifetime, the attack depleted the battery within only 83 hours. For the latter device, which is advertised with a one-month battery lifetime, the battery was exhausted within 13 hours. These results demonstrate that a single Diameter attack can drastically shorten the operational lifetime of battery-powered cellular devices from months or years to only a few days or even hours.


\vspace{-5pt}
\section{Generalizability Beyond Diameter}
\label{sec: generalizability beyond diameter}
 \vspace{-5pt}

Although \name\ is developed for Diameter exploitability analysis, its underlying methodology is not tied to Diameter. The core capability provided by \name\ is the inference of protocol-specific exploitability-analysis guidance from a small number of examples. We therefore ask whether this capability can generalize beyond Diameter and support exploitability analysis for protocols in other domains.

To answer this question, we apply \name\ to 16 additional widely used protocols spanning diverse domains, such as VoIP communication, Internet routing, and IoT messaging. Importantly, no component of \name\ is modified. \magentaadd{Using GPT-5.4 and} the same configuration used for Diameter analysis (Section~\ref{subsec: configuration study}), we simply replace the Diameter examples used for prompt inference with one exploitable example and one non-exploitable example specific to the target protocol, and then use the inferred guidance to analyze the target protocol. In total, the evaluated protocols comprise 857 protocol fields.
Table~\ref{table:generalizability} summarizes the results. Across the 857 protocol fields, \name\ generated 279 protocol-level PoCs, including 129 previously reported vulnerabilities and 150 previously unreported ones. To assess whether these findings affect real implementations, we further evaluated the 150 previously unreported ones on representative open-source implementations. We confirmed that 118 of them can successfully trigger their predicted security impacts, achieving attacks such as unauthorized message injection, session hijacking, response manipulation, denial of service, and information leakage.
Moreover, we have responsibly disclosed these vulnerabilities to the affected open-source projects. To date, 8 vulnerabilities affecting 7 vendors have been confirmed. The disclosures have resulted in 1 CVE assignment and 7 CNVD identifiers, while many additional reports remain under review.

Overall, these results suggest that \name\ does not rely on protocol-specific designs tailored to Diameter. Without modifying its methodology, \name\ remains effective across a diverse set of protocols, demonstrating the strong transferability of the methodology underlying \name.



\vspace{-5pt}
\section{Discussion}
\label{sec: discussion}
\vspace{-5pt}

\noindent
\textbf{\magentaadd{\name and protocol fuzzing.}}
\magentaadd{
One may ask how \name\ compares with protocol fuzzing. To examine this, we construct a Diameter fuzzing baseline on top of DiaFuzzer~\cite{orange_diafuzzer}, an open-source Diameter message fuzzer, and compare it with \name\ on the same 64 implementation-confirmed PoCs from Section~\ref{subsec: implementation-level findings}.
To focus the comparison on methodological differences rather than DiaFuzzer-specific limitations, we strengthen this baseline to systematically mutate each target field across all functional values explicitly defined in the Diameter specifications and add four security oracles following prior LTE fuzzing and dynamic-testing work~\cite{bennett2024ransacked, park2022doltest, kim2019touching} --- automated checks for LTE connectivity loss, data-service failure, voice-service failure, and information leakage.
Table~\ref{table: fuzzing comparison} in the appendix shows the results.
With these extensions, the fuzzing baseline generates the same field-value manipulations used in all 64 PoCs produced by \name, but automatically detects the security effects of only 26 (40.63\%).
For the remaining 38 PoCs, detecting their security effects requires additional test logic: for 29, the security effect becomes observable only after a specific subsequent action, such as UE reattachment, roaming, location change, TAU, or S-CSCF restart, while the other 9 require custom detection procedures tailored to their specific security effects (e.g., bearer QoS degradation, reduced data rate, battery drain, or IMS service failure). 
Thus, the fuzzing baseline fails to identify these 38 PoCs as successful attacks.
These results suggest that a stateful fuzzer could in principle identify these attacks if the necessary subsequent actions and security checks were known and encoded in advance. Determining them, however, requires reasoning about how each field affects downstream protocol behavior. This is precisely the field-level exploitability problem that \name\ is designed to address: by reasoning over field semantics and downstream behavior, \name\ identifies exploitable field-value manipulations, constructs the corresponding attack procedures, and predicts their security effects.
}

\noindent\textbf{From case-based discovery to systematic discovery}.
A longstanding challenge in security analysis is that many security discoveries rely on expert knowledge accumulated through experience, yet often difficult to explicitly articulate and transfer, making vulnerability discovery largely a manual and case-by-case process. Our results suggest a potential opportunity to change this dynamic. By inferring reusable security-analysis guidance from a small number of expert-discovered cases, \name provides evidence that such knowledge can be systematically captured, externalized, and propagated. More broadly, this raises the possibility that security discovery may gradually evolve from a predominantly case-based activity toward a more systematic process.

\vspace{2pt}\noindent\textbf{The challenge of practical mitigation}.
Given that the vulnerabilities discovered by \name\ rely on publicly reachable Diameter interfaces and disabled TLS protection, a natural question is whether the risks can be eliminated by simply addressing these two conditions. However, our discussions with operators and vendors suggest that publicly reachable core-network elements and disabled TLS protection will continue to exist in operational environments due to complex operational considerations, such as deployment, maintenance, interoperability, and performance requirements. As a result, new risks continue to be discovered and threaten core-network operations, while the insecure deployment conditions are unlikely to disappear anytime soon, creating a persistent operational challenge for core networks.

\vspace{2pt}\noindent\textbf{Limitation}.
One limitation of this work is the size of the manually labeled evaluation dataset introduced in Section~\ref{subsec: End-to-End Effectiveness}. Constructing ground-truth labels for Diameter exploitability analysis requires expert-driven protocol analysis and, in many cases, PoC construction, making large-scale labeling expensive and time-consuming. As a result, our evaluation is conducted on a dataset of 40 protocol fields. While this dataset covers both exploitable and non-exploitable fields across the S6a and Cx interfaces and the effectiveness of \name\ is further supported by large-scale protocol analysis and implementation-level validation, a larger ground-truth dataset would further strengthen the evaluation.

\vspace{-5pt}
\section{Related Work}
\label{sec:related}
\vspace{-5pt}

\noindent\textbf{Security analysis of cellular networks}.
Prior work has developed automated techniques for analyzing cellular-network security, mainly through testing-based analysis~\cite{chen2023sherlock,al2024hermes,park2022doltest,tu2024logic}, formal verification~\cite{basin2018formal,akon2023formal,akon2025control}, and fuzzing~\cite{dong2025corecrisis,klischies2025basebridge,bennett2024ransacked,hoang2025llfuzz}. Testing-based approaches typically execute cellular protocol procedures or implementations and analyze whether their behaviors follow expected protocol specifications. Formal methods usually construct models of protocol procedures and verify predefined security properties. Fuzzing-based approaches often mutate signaling messages~\cite{klischies2025basebridge}, protocol fields~\cite{bennett2024ransacked}, or protocol states~\cite{dong2025corecrisis} to expose abnormal implementation behavior. Collectively, these works primarily study whether cellular protocol procedures satisfy expected security properties, whether implementations comply with protocol requirements, or whether malformed signaling messages can expose implementation bugs. Our work studies a different problem: automatically determining whether protocol-compliant functionality can be leveraged by attackers to cause security consequences.

\vspace{2pt}\noindent\textbf{LLM-assisted security analysis.}
Recent work has applied LLMs to a wide range of security tasks, such as penetration testing~\cite{deng2024pentestgpt,incalmo2026,li2025iris}, fuzzing\cite{song2026protocolguard,ma2024one,xia2024fuzz4all}, vulnerability detection~\cite{ullah2024llms,lekssays2025llmxcpg}, and vulnerability exploitation~\cite{ji2026firmagent,meng2024chatafl,abramovich2024enigma}. PentestGPT~\cite{deng2024pentestgpt} and FirmAgent~\cite{ji2026firmagent} use LLMs as agents to orchestrate penetration testing of general systems and IoT firmware. ChatAFL~\cite{meng2024chatafl}, TitanFuzz~\cite{deng2023titanfuzz}, and Fuzz4All~\cite{xia2024fuzz4all} use LLMs to generate or mutate test inputs for fuzzing. IRIS~\cite{li2025iris} uses LLMs to infer taint specifications for static vulnerability detection, while EnIGMA~\cite{abramovich2024enigma} equips LLM agents with interactive tools for autonomous vulnerability exploitation.
Like these systems, \name\ also relies on task-specific information to guide security analysis. However, the information required by existing systems is typically explicit and available, such as source code, protocol grammars, execution traces, or tool outputs. In contrast, our analysis requires exploitability-analysis discipline and evidence-collection guidance, both of which are tacit expert knowledge that is not available in an explicit form. Our work therefore not only performs exploitability analysis, but must also address the problem of obtaining such knowledge to support the analysis.


\vspace{-5pt}
\section{Conclusion}
\vspace{-5pt}

In this paper, we presented \name, an LLM-based methodology for reliable field-level protocol exploitability analysis. Using \name, we conducted the first systematic field-level exploitability analysis of the Diameter S6a and Cx interfaces and found that \magentaadd{over one quarter} of the analyzed fields are exploitable once core-network elements become reachable. Our analysis further uncovered 95 previously unknown vulnerabilities in open-source and commercial EPC/IMS implementations, indicating that many of these protocol-level weaknesses are already present in existing implementations.
\vspace{-5pt}
\bibliographystyle{IEEEtran}
\bibliography{ref}

@misc{lens,
  author       = {LENS},
  title        = {{LENS: Artifact}},
  howpublished = {Online. Available: \url{https://github.com/HKUSL/LENS}},
  note         = {GitHub repository},
  year         = {2026}
}

@incollection{polanyi2009tacit,
  title={The tacit dimension},
  author={Polanyi, Michael},
  booktitle={Knowledge in organisations},
  pages={135--146},
  year={2009},
  publisher={Routledge}
}

@techreport{miller2026badconnection,
  author      = {Miller, Gary and Lange, Swantje},
  title       = {Bad Connection: Uncovering Global Telecom Exploitation by Covert Surveillance Actors},
  institution = {The Citizen Lab, University of Toronto},
  type        = {Citizen Lab Report},
  number      = {192},
  year        = {2026},
  month       = apr,
  url={https://citizenlab.ca/research/uncovering-global-telecom-exploitation-by-covert-surveillance-actors/},
  note        = {Published April 23, 2026}
}

@inproceedings{tan2021device,
  title={Device-centric detection and mitigation of diameter signaling attacks against mobile core},
  author={Tan, Zhaowei and Ding, Boyan and Zhang, Zhehui and Li, Qianru and Guo, Yunqi and Lu, Songwu},
  booktitle={2021 IEEE Conference on Communications and Network Security (CNS)},
  pages={29--37},
  year={2021},
  organization={IEEE}
}

@inproceedings{lin2025ado,
  title={Ado: Automatic data optimization for inputs in llm prompts},
  author={Lin, Sam and Hua, Wenyue and Li, Lingyao and Wang, Zhenting and Zhang, Yongfeng},
  booktitle={Findings of the Association for Computational Linguistics: ACL 2025},
  pages={26134--26146},
  year={2025}
}

@inproceedings{cai2024large,
  title={Large language models as tool makers},
  author={Cai, Tianle and Wang, Xuezhi and Ma, Tengyu and Chen, Xinyun and Zhou, Denny},
  booktitle={International Conference on Learning Representations},
  volume={2024},
  pages={54067--54089},
  year={2024}
}

@inproceedings{lekssays2025llmxcpg,
  title={$\{$LLMxCPG$\}$:$\{$Context-Aware$\}$ vulnerability detection through code property $\{$Graph-Guided$\}$ large language models},
  author={Lekssays, Ahmed and Mouhcine, Hamza and Tran, Khang and Yu, Ting and Khalil, Issa},
  booktitle={34th USENIX Security Symposium (USENIX Security 25)},
  pages={489--507},
  year={2025}
}

@inproceedings{ullah2024llms,
  title={Llms cannot reliably identify and reason about security vulnerabilities (yet?): A comprehensive evaluation, framework, and benchmarks},
  author={Ullah, Saad and Han, Mingji and Pujar, Saurabh and Pearce, Hammond and Coskun, Ayse and Stringhini, Gianluca},
  booktitle={2024 IEEE symposium on security and privacy (SP)},
  pages={862--880},
  year={2024},
  organization={IEEE}
}

@inproceedings{honovich2023instruction,
  title={Instruction induction: From few examples to natural language task descriptions},
  author={Honovich, Or and Shaham, Uri and Bowman, Samuel and Levy, Omer},
  booktitle={Proceedings of the 61st Annual Meeting of the Association for Computational Linguistics (Volume 1: Long Papers)},
  pages={1935--1952},
  year={2023}
}

@techreport{rfc6733,
  author      = {Fajardo, Victor and Arkko, Jari and Loughney, John and Zorn, Glen},
  title       = {Diameter Base Protocol},
  institution = {Internet Engineering Task Force (IETF)},
  type        = {RFC},
  number      = {6733},
  year        = {2012},
  month       = oct
}

@techreport{3GPP_TS_24301_T3412,
  author      = {{3GPP}},
  title       = {{Non-Access-Stratum (NAS) protocol for Evolved Packet System (EPS); Stage 3}},
  institution = {{3rd Generation Partnership Project (3GPP)}},
  type        = {{Technical Specification}},
  number      = {{TS 24.301}},
  year        = {2023}
}

@techreport{3GPP_TS_23401,
  author      = {{3GPP}},
  title       = {{General Packet Radio Service (GPRS) enhancements for Evolved Universal Terrestrial Radio Access Network (E-UTRAN) access}},
  institution = {{3rd Generation Partnership Project (3GPP)}},
  type        = {{Technical Specification}},
  number      = {{TS 23.401}},
  year      = {2026}
}

@techreport{29230,
  author      = {3GPP},
  title       = {Diameter applications; 3GPP specific codes and identifiers},
  institution = {3rd Generation Partnership Project (3GPP)},
  type        = {Technical Specification},
  number      = {TS 29.230, Version 19.3.0, Release 19},
  year        = {2026},
  month       = jan,
}

@techreport{29272,
  author      = {3GPP},
  title       = {Evolved Packet System (EPS); Mobility Management Entity (MME) and Serving GPRS Support Node (SGSN) related interfaces based on Diameter protocol},
  institution = {3rd Generation Partnership Project (3GPP)},
  type        = {Technical Specification},
  number      = {TS 29.272, Version 19.4.0, Release 19},
    year      = {2026}
}

@inproceedings{yang2026smallcell,
  author    = {Yang, Yaru and Zhang, Yiming and Wan, Tao and Duan, Haixin and Chang, Deliang and Li, Yishen and Tang, Shujun},
  title     = {Small Cell, Big Risk: A Security Assessment of 4G LTE Femtocells in the Wild},
  booktitle = {Proceedings of the Network and Distributed System Security Symposium (NDSS)},
  year      = {2026},
  doi       = {10.14722/ndss.2026.241968}
}

@techreport{29228,
  author      = {3GPP},
  title       = {IP Multimedia (IM) Subsystem Cx and Dx Interfaces; Signalling flows and message contents},
  institution = {3rd Generation Partnership Project (3GPP)},
  type        = {Technical Specification},
  number      = {TS 29.228, Version 19.2.0, Release 19},
  year        = {2026},
  month       = jan
}

@techreport{29229,
  author      = {3GPP},
  title       = {Cx and Dx interfaces based on the Diameter protocol; Protocol details},
  institution = {3rd Generation Partnership Project (3GPP)},
  type        = {Technical Specification},
  number      = {TS 29.229, Version 19.1.0, Release 19},
  year        = {2026}
}

@article{kwon2021towards,
  title={Towards 5G-based IoT security analysis against Vo5G eavesdropping},
  author={Kwon, Sungmoon and Park, Seongmin and Cho, HyungJin and Park, Youngkwon and Kim, Dowon and Yim, Kangbin},
  journal={Computing},
  volume={103},
  number={3},
  pages={425--447},
  year={2021},
  publisher={Springer}
}

@phdthesis{cui2025security,
  author = {Cui, Zhiwei},
  title  = {Research on Security Vulnerability Analysis Techniques for Key Applications in 5G Networks},
  school = {Beijing University of Posts and Telecommunications},
  year   = {2025},
  month  = aug,
  type   = {Doctoral Dissertation},
  url ={https://doi.org/10.26969/d.cnki.gbydu.2025.000345}
}

@inproceedings{zhang2025invade,
  title={Invade the walled garden: Evaluating gtp security in cellular networks},
  author={Zhang, Yiming and Wan, Tao and Yang, Yaru and Duan, Haixin and Wang, Yichen and Chen, Jianjun and Wei, Zixiang and Li, Xiang},
  booktitle={2025 IEEE Symposium on Security and Privacy (SP)},
  pages={1159--1177},
  year={2025},
  organization={IEEE}
}

@misc{kotte2016analysis,
  title={Analysis and Experimental Verification of Diameter Attacks in Long Term Evolution Networks},
  author={Kotte, Bhanu Teja},
  year={2016}
}

@article{mashukov2017diameter,
  title={Diameter Security: An Auditor's Viewpoint},
  author={Mashukov, Sergey},
  journal={Journal of ICT Standardization},
  volume={5},
  number={1},
  pages={53--68},
  year={2017},
  publisher={River Publishers}
}

@article{holtmanns2018mobile,
  title={Mobile subscriber profile data privacy breach via 4G diameter interconnection},
  author={Holtmanns, Silke and Oliver, Ian and Miche, Yoan},
  journal={Journal of ICT Standardization},
  volume={6},
  number={3},
  pages={245--262},
  year={2018},
  publisher={River Publishers}
}

@inproceedings{al2024hermes,
  title={Hermes: Unlocking security analysis of cellular network protocols by synthesizing finite state machines from natural language specifications},
  author={Al Ishtiaq, Abdullah and Das, Sarkar Snigdha Sarathi and Rashid, Syed Md Mukit and Ranjbar, Ali and Tu, Kai and Wu, Tianwei and Song, Zhezheng and Wang, Weixuan and Akon, Mujtahid and Zhang, Rui and others},
  booktitle={33rd USENIX Security Symposium (USENIX Security 24)},
  pages={4445--4462},
  year={2024}
}

@inproceedings{dong2025corecrisis,
  title={CoreCrisis: Threat-Guided and Context-Aware Iterative Learning and Fuzzing of 5G Core Networks},
  author={Dong, Yilu and Yang, Tianchang and Al Ishtiaq, Abdullah and Rashid, Syed Md Mukit and Ranjbar, Ali and Tu, Kai and Wu, Tianwei and Mahmud, Md Sultan and Hussain, Syed Rafiul},
  booktitle={34th USENIX Security Symposium (USENIX Security 25)},
  pages={5287--5306},
  year={2025}
}

@inproceedings{klischies2025basebridge,
  title={Basebridge: Bridging the gap between over-the-air and emulation testing for cellular baseband firmware},
  author={Klischies, Daniel and Goos, Dyon and Hirsch, David and Milburn, Alyssa and Muench, Marius and Moonsamy, Veelasha},
  booktitle={2025 IEEE Symposium on Security and Privacy (SP)},
  pages={1101--1119},
  year={2025},
  organization={IEEE}
}

@inproceedings{chen2023sherlock,
  title={Sherlock on specs: Building $\{$LTE$\}$ conformance tests through automated reasoning},
  author={Chen, Yi and Tang, Di and Yao, Yepeng and Zha, Mingming and Wang, XiaoFeng and Liu, Xiaozhong and Tang, Haixu and Liu, Baoxu},
  booktitle={32nd USENIX Security Symposium (USENIX Security 23)},
  pages={3529--3545},
  year={2023}
}

@inproceedings{akon2023formal,
  title={Formal analysis of access control mechanism of 5G core network},
  author={Akon, Mujtahid and Yang, Tianchang and Dong, Yilu and Hussain, Syed Rafiul},
  booktitle={Proceedings of the 2023 ACM SIGSAC conference on computer and communications security},
  pages={666--680},
  year={2023}
}

@inproceedings{bennett2024ransacked,
  title={Ransacked: A domain-informed approach for fuzzing lte and 5g ran-core interfaces},
  author={Bennett, Nathaniel and Zhu, Weidong and Simon, Benjamin and Kennedy, Ryon and Enck, William and Traynor, Patrick and Butler, Kevin RB},
  booktitle={Proceedings of the 2024 on ACM SIGSAC Conference on Computer and Communications Security},
  pages={2027--2041},
  year={2024}
}

@article{gao2023retrieval,
  title={Retrieval-augmented generation for large language models: A survey},
  author={Gao, Yunfan and Xiong, Yun and Gao, Xinyu and Jia, Kangxiang and Pan, Jinliu and Bi, Yuxi and Dai, Yixin and Sun, Jiawei and Wang, Haofen and Wang, Haofen and others},
  journal={arXiv preprint arXiv:2312.10997},
  volume={2},
  number={1},
  pages={32},
  year={2023}
}

@techreport{3GPP_TS_23003,
  author      = {3GPP},
  title       = {Numbering, addressing and identification},
  institution = {3rd Generation Partnership Project (3GPP)},
  type        = {Technical Specification},
  number      = {TS 23.003},
  version     = {19.6.0},
  year        = {2026},
  month       = mar
}

@inproceedings{hoang2025llfuzz,
  title={LLFUZZ: An Over-the-Air Dynamic Testing Framework for Cellular Baseband Lower Layers},
  author={Hoang, Tuan Dinh and Oh, Taekkyung and Park, CheolJun and Yun, Insu and Kim, Yongdae},
  booktitle={34th USENIX Security Symposium (USENIX Security 25)},
  pages={5267--5286},
  year={2025},
  organization={USENIX Association}
}

@inproceedings{park2022doltest,
  title={$\{$DoLTEst$\}$: In-depth downlink negative testing framework for $\{$LTE$\}$ devices},
  author={Park, CheolJun and Bae, Sangwook and Oh, BeomSeok and Lee, Jiho and Lee, Eunkyu and Yun, Insu and Kim, Yongdae},
  booktitle={31st USENIX Security Symposium (USENIX Security 22)},
  pages={1325--1342},
  year={2022}
}

@inproceedings{tu2024logic,
  title={Logic gone astray: A security analysis framework for the control plane protocols of 5g basebands},
  author={Tu, Kai and Al Ishtiaq, Abdullah and Rashid, Syed Md Mukit and Dong, Yilu and Wang, Weixuan and Wu, Tianwei and Hussain, Syed Rafiul},
  booktitle={33rd USENIX Security Symposium (USENIX Security 24)},
  pages={3063--3080},
  year={2024}
}

@inproceedings{ma2024one,
  title = {From One Thousand Pages of Specification to Unveiling Hidden Bugs: Large Language Model Assisted Fuzzing of Matter {IoT} Devices},
  author = {Ma, Xiaoyue and Luo, Lannan and Zeng, Qiang},
  booktitle = {33rd USENIX Security Symposium (USENIX Security 24)},
  pages = {4783--4800},
  year = {2024},
  publisher = {USENIX Association}
}

@inproceedings{song2026protocolguard,
  title = {ProtocolGuard: Detecting Protocol Non-compliance Bugs via {LLM}-guided Static Analysis and Dynamic Verification},
  author = {Song, Xiangpu and Pei, Longjia and Wu, Jianliang and Zeng, Yingpei and He, Gaoshuo and Zuo, Chaoshun and Liu, Xiaofeng and Zhao, Qingchuan and Guo, Shanqing},
  booktitle = {Proceedings 2026 Network and Distributed System Security Symposium (NDSS)},
  year = {2026},
  publisher = {Internet Society},
  doi = {10.14722/ndss.2026.240521}
}

@inproceedings{deng2024pentestgpt,
  title = {{PentestGPT}: Evaluating and Harnessing Large Language Models for Automated Penetration Testing},
  author = {Deng, Gelei and Liu, Yi and Mayoral-Vilches, V{\'\i}ctor and Liu, Peng and Li, Yuekang and Xu, Yuan and Zhang, Tianwei and Liu, Yang and Pinzger, Martin and Rass, Stefan},
  booktitle = {33rd USENIX Security Symposium (USENIX Security 24)},
  pages = {847--864},
  year = {2024},
  publisher = {USENIX Association}
}

@inproceedings{incalmo2026,
  title = {Incalmo: An Autonomous {LLM}-assisted System for Red Teaming Multi-Host Networks},
  author = {Singer, Brian and Lucas, Keane and Adiga, Lakshmi and Jain, Meghna and Bauer, Lujo and Sekar, Vyas},
  booktitle = {2026 IEEE Symposium on Security and Privacy (SP)},
  year = {2026},
  note = {To appear}
}

@article{yu2024auto,
  title={Auto-rag: Autonomous retrieval-augmented generation for large language models},
  author={Yu, Tian and Zhang, Shaolei and Feng, Yang},
  journal={arXiv preprint arXiv:2411.19443},
  year={2024}
}

@inproceedings{wang2025llms,
  title={Llms know what they need: Leveraging a missing information guided framework to empower retrieval-augmented generation},
  author={Wang, Keheng and Duan, Feiyu and Li, Peiguang and Wang, Sirui and Cai, Xunliang},
  booktitle={Proceedings of the 31st International Conference on Computational Linguistics},
  pages={2379--2400},
  year={2025}
}

@article{xu2026beyond,
  title={Beyond Elicitation: Provision-based Prompt Optimization for Knowledge-Intensive Tasks},
  author={Xu, Yunzhe and Zhang, Zhuosheng and Liu, Zhe},
  journal={IEEE Transactions on Audio, Speech and Language Processing},
  year={2026},
  publisher={IEEE}
}

@inproceedings{basin2018formal,
  title={A formal analysis of 5G authentication},
  author={Basin, David and Dreier, Jannik and Hirschi, Lucca and Radomirovic, Sa{\v{s}a and Sasse, Ralf and Stettler, Vincent}},
  booktitle={Proceedings of the 2018 ACM SIGSAC conference on computer and communications security},
  pages={1383--1396},
  year={2018}
}

@inproceedings{akon2025control,
  title={From control to chaos: A comprehensive formal analysis of 5g's access control},
  author={Akon, Mujtahid and Toufikuzzaman, Md and Hussain, Syed Rafiul},
  booktitle={2025 IEEE Symposium on Security and Privacy (SP)},
  pages={1081--1100},
  year={2025},
  organization={IEEE}
}

@misc{open5gs2026,
  author       = {S. Lee},
  title        = {Open5GS: Open Source implementation for 5G Core and EPC},
  howpublished = {[Online]. Available: https://github.com/open5gs/open5gs},
  year         = {2026},
  note         = {Version v2.7.7}
}

@misc{kamailio2026,
  author       = {Kamailio SIP Server Project},
  title        = {Kamailio: The Open Source SIP Server},
  howpublished = {[Online]. Available: https://github.com/kamailio/kamailio},
  year         = {2026},
  note         = {Version v6.1.3}
}

@article{abramovich2024enigma,
  title = {{EnIGMA}: Interactive Tools Substantially Assist {LM} Agents in Finding Security Vulnerabilities},
  author = {Abramovich, Talor and Udeshi, Meet and Shao, Minghao and Lieret, Kilian and Xi, Haoran and Milner, Kimberly and Jancheska, Sofija and Yang, John and Jimenez, Carlos E. and Khorrami, Farshad and others},
  journal = {arXiv preprint arXiv:2409.16165},
  year = {2024}
}

@inproceedings{deng2023titanfuzz,
  title = {Large Language Models are Zero-Shot Fuzzers: Fuzzing Deep-Learning Libraries via Large Language Models},
  author = {Deng, Yinlin and Xia, Chunqiu Steven and Peng, Haoran and Yang, Chenyuan and Zhang, Lingming},
  booktitle = {Proceedings of the 32nd ACM SIGSOFT International Symposium on Software Testing and Analysis (ISSTA)},
  pages = {423--435},
  year = {2023},
  publisher = {ACM}
}

@inproceedings{ji2026firmagent,
  title = {{FirmAgent}: Leveraging Fuzzing to Assist {LLM} Agents with {IoT} Firmware Vulnerability Discovery},
  author = {Ji, Jiangan and Zhang, Chao and Gan, Shuitao and Jian, Lin and Liu, Hangtian and Liu, Tieming and Zheng, Lei and Jia, Zhipeng},
  booktitle = {33rd Annual Network and Distributed System Security Symposium (NDSS)},
  year = {2026},
  publisher = {The Internet Society}
}

@inproceedings{li2025iris,
  title = {{IRIS}: {LLM}-Assisted Static Analysis for Detecting Security Vulnerabilities},
  author = {Li, Ziyang and Dutta, Saikat and Naik, Mayur},
  booktitle = {International Conference on Learning Representations (ICLR)},
  year = {2025}
}

@inproceedings{meng2024chatafl,
  title = {Large Language Model Guided Protocol Fuzzing},
  author = {Meng, Ruijie and Mirchev, Martin and B{\"o}hme, Marcel and Roychoudhury, Abhik},
  booktitle = {31st Annual Network and Distributed System Security Symposium (NDSS)},
  year = {2024},
  publisher = {The Internet Society}
}

@inproceedings{xia2024fuzz4all,
  title = {{Fuzz4All}: Universal Fuzzing with Large Language Models},
  author = {Xia, Chunqiu Steven and Paltenghi, Matteo and Tian, Jia Le and Pradel, Michael and Zhang, Lingming},
  booktitle = {Proceedings of the 46th IEEE/ACM International Conference on Software Engineering (ICSE)},
  pages = {126:1--126:13},
  year = {2024},
  publisher = {ACM}
}

@inproceedings{kim2019touching,
  title={Touching the untouchables: Dynamic security analysis of the LTE control plane},
  author={Kim, Hongil and Lee, Jiho and Lee, Eunkyu and Kim, Yongdae},
  booktitle={2019 IEEE Symposium on Security and Privacy (SP)},
  pages={1153--1168},
  year={2019},
  organization={IEEE}
}

@misc{orange_diafuzzer,
  author       = {{Orange OpenSource}},
  title        = {{diafuzzer}: Diameter Fuzzer},
  howpublished = {\url{https://github.com/Orange-OpenSource/diafuzzer}},
}
\vspace{-5pt}

\appendices
\vspace{-5pt}
\section{Open Science}
\label{sec:open-science}
\vspace{-5pt}

We publicly release our artifact, including the source code, all prompts, evaluation data, and protocol-level PoCs, \url{https://github.com/HKUSL/LENS}.
\vspace{-5pt}

\section{Acknowledgment}
\label{sec:acknowledgment}
\vspace{-5pt}

This work was supported by the National Natural Science Foundation of China under Grant U23B2024, the Xi’an Young and Middle-aged Leading Talents Program under Grant 25ZQRC00021, and OPPO.

\vspace{-5pt}
\section{Additional Results}
\label{sec:appendix-results}
\vspace{-5pt}

\begin{table}[H]
\centering
\caption{End-to-End effectiveness of \name{}.}
\label{table:appendix-end-to-end-s6a-cx}

\footnotesize
\setlength{\tabcolsep}{3pt}
\renewcommand{\arraystretch}{1.12}

\begin{tabular*}{\columnwidth}
{@{\hspace{\tabcolsep}\extracolsep{\fill}}lcrrr@{\hspace{\tabcolsep}}}
\toprule

\textbf{Model}
& \textbf{Interface}
& \multicolumn{1}{c}{\textbf{Recall (\%)}}
& \multicolumn{1}{c}{\textbf{Precision (\%)}}
& \multicolumn{1}{c}{\textbf{Accuracy (\%)}} \\

\midrule

\multirow{2}{*}{GPT}
& S6a
& 94.00\textpm{}8.94
& 90.63\textpm{}6.01
& 92.00\textpm{}5.70 \\
& Cx
& 98.00\textpm{}4.47
& 87.73\textpm{}7.11
& 92.00\textpm{}6.71 \\

\midrule

\multirow{2}{*}{GLM}
& S6a
& 94.00\textpm{}5.48
& 69.83\textpm{}7.96
& 76.00\textpm{}7.42 \\
& Cx
& 94.00\textpm{}5.48
& 90.55\textpm{}6.45
& 92.00\textpm{}5.70 \\

\midrule

\multirow{2}{*}{Claude}
& S6a
& 96.00\textpm{}5.48
& 83.08\textpm{}4.68
& 88.00\textpm{}2.74 \\
& Cx
& 100.00\textpm{}0.00
& 89.93\textpm{}8.27
& 94.00\textpm{}5.48 \\

\midrule

\multirow{2}{*}{Gemini}
& S6a
& 98.00\textpm{}4.47
& 77.82\textpm{}3.19
& 85.00\textpm{}3.54 \\
& Cx
& 94.00\textpm{}5.48
& 93.38\textpm{}10.17
& 93.00\textpm{}5.70 \\

\bottomrule
\end{tabular*}
\vspace{-5pt}
\end{table}






\begin{table}[H]
\centering
\caption{Per-field correctness of \name{} across five runs.}
\label{table:per-field-correctness}

\footnotesize
\setlength{\tabcolsep}{3.5pt}
\renewcommand{\arraystretch}{1.15}

\begin{tabular*}{\columnwidth}
{@{\hspace{\tabcolsep}\extracolsep{\fill}}lrrrrrr@{\hspace{\tabcolsep}}}
\toprule
\textbf{Model}
& \multicolumn{1}{c}{\textbf{5/5}}
& \multicolumn{1}{c}{\textbf{4/5}}
& \multicolumn{1}{c}{\textbf{3/5}}
& \multicolumn{1}{c}{\textbf{2/5}}
& \multicolumn{1}{c}{\textbf{1/5}}
& \multicolumn{1}{c}{\textbf{0/5}} \\
\midrule
GPT    & 30 & 7 & 2 & 0 & 0 & 1 \\
GLM    & 29 & 2 & 3 & 2 & 2 & 2 \\
Claude & 31 & 4 & 2 & 2 & 1 & 0 \\
Gemini & 29 & 6 & 2 & 1 & 1 & 1 \\
\bottomrule
\end{tabular*}

\vspace{5pt}

{\footnotesize
\parbox{\columnwidth}{
\textit{Note:} Each column reports the number of fields correctly analyzed
in exactly $k$ of five runs (e.g., 5/5 means correct in all five runs).
}}

\end{table}
\begin{table*}[p]
\centering
\caption{Message-level exploitability results.}
\label{table: message-level findings}

\fontsize{7.2pt}{8.4pt}\selectfont
\setlength{\tabcolsep}{3.2pt}
\renewcommand{\arraystretch}{1.18}
\setlength{\arrayrulewidth}{0.25pt}
\setlength{\extrarowheight}{0.6pt}

\newlength{\digitwidth}
\settowidth{\digitwidth}{0}

\newcommand{\nocell}[1]{%
  \makebox[0pt][l]{%
    \hspace{\dimexpr\digitwidth/2\relax}%
    \makebox[0pt][r]{#1}%
  }%
}

{
\resizebox{\textwidth}{!}{%
\begin{tabular}{|c|c|c|c|l|r|r|r|}
\hline

\textbf{No.}
& \textbf{Interface}
& \textbf{Message}
& \textbf{Receiver}
& \textbf{Receiver-Side Operation}
& \multicolumn{1}{c|}{\textbf{Fields}}
& \multicolumn{1}{c|}{\textbf{PoCs}}
& \textbf{Effect Categories} \\

\hline

\nocell{1}
& S6a
& ULR
& MME
& Report supported feature information for location update
& 1
& 1
& C3 \\
\hline

\nocell{2}
& S6a
& AIR
& MME
& Report supported feature information for S6a authentication
& 1
& 1
& C3 \\
\hline

\nocell{3}
& S6a
& NOR
& MME
& Report supported feature information for S6a notification
& 1
& 1
& C3 \\
\hline

\nocell{4}
& S6a
& PUR
& MME
& Report supported feature information for UE purge
& 1
& 1
& C3 \\
\hline

\nocell{5}
& S6a
& IDR
& MME
& Report the UE's current state
& 2
& 3
& C5 \\
\hline

\nocell{6}
& S6a
& IDR
& MME
& Report the UE's location information
& 4
& 4
& C5 \\
\hline

\nocell{7}
& S6a
& DSR
& MME
& Update selected subscriber/service data
& 6
& 8
& C1, C2 \\
\hline

\nocell{8}
& S6a
& IDR
& MME
& Update the UE's aggregate data-rate limit
& 6
& 8
& C2 \\
\hline

\nocell{9}
& S6a
& IDR
& MME
& Update the UE's APN/PDN connectivity configuration
& 18
& 21
& C1, C2 \\
\hline

\nocell{10}
& S6a
& RSR
& MME
& Update the UE's APN/PDN connectivity configuration
& 3
& 4
& C1, C2 \\
\hline

\nocell{11}
& S6a
& IDR
& MME
& Update the UE's CSG connectivity configuration
& 2
& 2
& C1 \\
\hline

\nocell{12}
& S6a
& IDR
& MME
& Update the UE's LCS configuration
& 11
& 11
& C2 \\
\hline

\nocell{13}
& S6a
& RSR
& MME
& Update the UE's LCS configuration
& 2
& 2
& C2 \\
\hline

\nocell{14}
& S6a
& IDR
& MME
& Update the UE's MDT configuration
& 27
& 27
& C2, C4, C5 \\
\hline

\nocell{15}
& S6a
& RSR
& MME
& Update the UE's MDT configuration
& 15
& 15
& C2, C4, C5 \\
\hline

\nocell{16}
& S6a
& IDR
& MME
& Update the UE's ProSe configuration
& 3
& 3
& C2 \\
\hline

\nocell{17}
& S6a
& RSR
& MME
& Update the UE's ProSe configuration
& 2
& 2
& C2 \\
\hline

\nocell{18}
& S6a
& CLR
& MME
& Update the UE's registration state
& 1
& 2
& C1 \\
\hline

\nocell{19}
& S6a
& IDR
& MME
& Update the UE's subscribed QoS profile
& 4
& 4
& C2 \\
\hline

\nocell{20}
& S6a
& IDR
& MME
& Update the UE's subscriber profile
& 23
& 23
& C1, C2, C4 \\
\hline

\nocell{21}
& S6a
& RSR
& MME
& Update the UE's subscriber profile
& 10
& 10
& C1, C2, C4 \\
\hline

\nocell{22}
& S6a
& IDR
& MME
& Update the UE's WLAN offload permission
& 2
& 2
& C1 \\
\hline

\nocell{23}
& S6a
& CLR
& HSS
& Report supported feature information for location cancellation
& 1
& 1
& C3 \\
\hline

\nocell{24}
& S6a
& RSR
& HSS
& Report supported feature information for S6a reset
& 1
& 1
& C3 \\
\hline

\nocell{25}
& S6a
& DSR
& HSS
& Report supported feature information for subscriber-data deletion
& 1
& 1
& C3 \\
\hline

\nocell{26}
& S6a
& IDR
& HSS
& Report supported feature information for subscriber-data insertion
& 1
& 1
& C3 \\
\hline

\nocell{27}
& S6a
& AIR
& HSS
& Report the UE's authentication vectors
& 4
& 5
& C1, C3 \\
\hline

\nocell{28}
& S6a
& ULR
& HSS
& Report the UE's subscriber profile
& 2
& 2
& C3, C5 \\
\hline

\nocell{29}
& S6a
& PUR
& HSS
& Update the UE's purged status
& 1
& 1
& C2 \\
\hline

\nocell{30}
& S6a
& NOR
& HSS
& Update the UE's reachability information
& 1
& 1
& C4 \\
\hline

\nocell{31}
& S6a
& ULR
& HSS
& Update the UE's serving-MME information
& 2
& 3
& C1, C3 \\
\hline

\nocell{32}
& S6a
& NOR
& HSS
& Update the UE-to-PDN-GW association
& 1
& 1
& C1 \\
\hline

\nocell{33}
& Cx
& MAR
& S-CSCF
& Report supported feature information for IMS authentication
& 1
& 1
& C3 \\
\hline

\nocell{34}
& Cx
& UAR
& S-CSCF
& Report supported feature information for IMS authorization
& 1
& 1
& C3 \\
\hline

\nocell{35}
& Cx
& SAR
& S-CSCF
& Report supported feature information for server assignment
& 1
& 1
& C3 \\
\hline

\nocell{36}
& Cx
& LIR
& S-CSCF
& Report supported feature information for user-location query
& 1
& 1
& C3 \\
\hline

\nocell{37}
& Cx
& PPR
& S-CSCF
& Update the UE's IMS charging configuration
& 2
& 2
& C1 \\
\hline

\nocell{38}
& Cx
& PPR
& S-CSCF
& Update the UE's IMS authentication data
& 9
& 9
& C1 \\
\hline

\nocell{39}
& Cx
& RTR
& S-CSCF
& Update the UE's IMS registration state
& 6
& 6
& C1, C3 \\
\hline

\nocell{40}
& Cx
& PPR
& S-CSCF
& Update the UE's IMS service profile
& 2
& 2
& C1 \\
\hline

\nocell{41}
& Cx
& PPR
& S-CSCF
& Update the UE's WebRTC configuration
& 2
& 2
& C1 \\
\hline

\nocell{42}
& Cx
& PPR
& HSS
& Report supported feature information for profile update
& 1
& 1
& C3 \\
\hline

\nocell{43}
& Cx
& RTR
& HSS
& Report supported feature information for registration termination
& 1
& 1
& C3 \\
\hline

\nocell{44}
& Cx
& MAR
& HSS
& Report the UE's IMS authentication vectors
& 5
& 5
& C1, C3 \\
\hline

\nocell{45}
& Cx
& UAR
& HSS
& Report the UE's IMS authorization result
& 2
& 3
& C1, C3 \\
\hline

\nocell{46}
& Cx
& SAR
& HSS
& Report the UE's IMS service profile
& 1
& 1
& C3 \\
\hline

\nocell{47}
& Cx
& LIR
& HSS
& Report the UE's serving S-CSCF information
& 2
& 3
& C3, C5 \\
\hline

\nocell{48}
& Cx
& SAR
& HSS
& Update the UE's restoration information
& 2
& 2
& C1 \\
\hline

\nocell{49}
& Cx
& SAR
& HSS
& Update the UE's serving S-CSCF information
& 5
& 9
& C1, C3 \\
\hline

\textbf{Overall}
& ---
& ---
& ---
& \textbf{49 message-level scenarios}
& \textbf{204}
& \textbf{222}
& --- \\

\hline
\end{tabular}%
}
}

\vspace{5pt}

{\footnotesize
\parbox{\textwidth}{
\textit{Note:} This table provides the message-level view discussed in
Section~\ref{subsec: protocol-level findings}. The 204 validated fields are grouped
by Diameter message, receiver, and receiver-side operation. Fields sharing
these three properties constitute the same message-level scenario, yielding
49 scenarios in total. Fields and PoCs denote the numbers of validated
exploitable fields and corresponding protocol-level PoCs, respectively.
Effect categories C1--C5 follow
Table~\ref{table: security consequences}.
}}

\vspace{-0.4em}
\end{table*}

\setcounter{dbltopnumber}{3}
\renewcommand{\dbltopfraction}{0.99}
\renewcommand{\textfraction}{0.01}

\setlength{\dblfloatsep}{14pt}
\setlength{\dbltextfloatsep}{8pt}

\newcommand{\fullcirc}{\raisebox{0.05ex}{\scalebox{1.45}{\ensuremath{\bullet}}}}
\newcommand{\emptycirc}{\raisebox{0.05ex}{\scalebox{1.45}{\ensuremath{\circ}}}}

\newcommand{\val}[1]{%
  \lstinline[basicstyle=\normalfont\footnotesize\itshape,breaklines=true,breakatwhitespace=false,columns=fullflexible,keepspaces=true]|#1|%
}

\newcommand{\valmid}[1]{%
  \raisebox{-0.65ex}{%
    \lstinline[basicstyle=\normalfont\footnotesize\itshape,breaklines=true,breakatwhitespace=false,columns=fullflexible,keepspaces=true]|#1|%
  }%
}

\newcommand{\midcell}[1]{%
  \raisebox{-0.65ex}{#1}%
}

\newcommand{\affectedimpl}{\begin{tabular}[c]{@{}c@{}}\textbf{Affected}\tabularnewline\textbf{Impl.}\end{tabular}}

\newcommand{\apnfield}[1]{%
  \begin{tabular}[c]{@{}l@{}}%
  IDR: Subscription-Data.APN-Configuration-Profile.APN-Configuration.\\%
  #1%
  \end{tabular}%
}

\newcolumntype{N}{>{\centering\arraybackslash}m{0.45cm}}
\newcolumntype{C}{>{\centering\arraybackslash}m{0.55cm}}
\newcolumntype{Y}{>{\centering\arraybackslash}m{0.38cm}}

\newcolumntype{P}{>{\hsize=1.2\hsize\linewidth=\hsize\raggedright\arraybackslash}X}
\newcolumntype{Q}{>{\hsize=0.8\hsize\linewidth=\hsize\raggedright\arraybackslash}X}

\newcolumntype{R}{>{\hsize=0.7\hsize\linewidth=\hsize\raggedright\arraybackslash}X}
\newcolumntype{S}{>{\hsize=1.3\hsize\linewidth=\hsize\raggedright\arraybackslash}X}

\begin{table*}[!t]
\centering
\caption{Detailed results of the implementation-level assessment.}
\vspace{-3pt}
\label{table: affected-diameter-fields}
\footnotesize
\setlength{\tabcolsep}{2pt}
\renewcommand{\arraystretch}{1.00}
{\footnotesize\itshape
For CID, C1: service disruption; C2: service degradation; C3: information leakage; C4: battery depletion; C5: subscriber tracking;\par
For Affected Impl., Impl.: implementation; O1: Open5GS; O2: Kamailio; C1: Commercial A; C2: Commercial B;
\fullcirc: affected; \emptycirc: not affected.\par}
\vspace{2pt}

\begin{subtable}[t]{\textwidth}
\centering
\vspace{-3pt}
\caption{S6a Interface}
\label{table: affected-diameter-fields-s6a}

{\renewcommand{\tabularxcolumn}[1]{m{#1}}
\begin{tabularx}{\textwidth}{|N|P|Q|C|Y|Y|}
\hline
\multirow{2}{*}{\textbf{PID}} &
\multirow{2}{*}{\textbf{Diameter Fields}} &
\multirow{2}{*}{\normalfont\textbf{Value}} &
\multirow{2}{*}{\textbf{CID}} &
\multicolumn{2}{|c|}{\affectedimpl} \tabularnewline
\cline{5-6}
&  &  &  & \textbf{O1} & \textbf{C1} \tabularnewline
\hline
1  & CLR: Cancellation-Type & \val{MME_UPDATE_PROCEDURE} & C1 & \fullcirc & \fullcirc \tabularnewline \hline
2  & CLR: Cancellation-Type & \val{SUBSCRIPTION_WITHDRAWAL} & C1 & \fullcirc & \fullcirc \tabularnewline \hline
3  & ULR: ULR-Flags & \val{Initial-Attach-Indicator} & C1 & \fullcirc & \fullcirc \tabularnewline \hline
4  & ULR: ULR-Flags & \val{Skip Subscriber Data} & C3 & \fullcirc & \fullcirc \tabularnewline \hline
5  & AIR: Requested-EUTRAN-Authentication-Info & \val{1} & C3 & \fullcirc & \fullcirc \tabularnewline \hline
6  & IDR: IDR-Flags & \val{EPS User State Request} & C5 & \fullcirc & \fullcirc \tabularnewline \hline
7  & IDR: IDR-Flags & \val{EPS Location Information Request} & C5 & \fullcirc & \emptycirc \tabularnewline \hline
8  & DSR: Context-Identifier & \val{Non-Default APN} & C1 & \emptycirc & \fullcirc \tabularnewline \hline
9  & DSR: DSR-Flags & \val{PDN subscription contexts Withdrawal} & C1 & \emptycirc & \fullcirc \tabularnewline \hline
10 & DSR: DSR-Flags & \val{MSISDN Withdrawal} & C1 & \emptycirc & \fullcirc \tabularnewline \hline
11 & IDR: Subscription-Data.APN-OI-Replacement & \val{<Attacker>.mnc<MNC>.mcc<MCC>.gprs} & C1 & \emptycirc & \fullcirc \tabularnewline \hline
12 & IDR: Subscription-Data.APN-Configuration-Profile & \val{Null} & C1 & \emptycirc & \fullcirc \tabularnewline \hline
13 & \apnfield{All-APN-Configurations-Included-Indicator} & \valmid{All_APN_CONFIGURATIONS_INCLUDED} & \midcell{C1} & \midcell{\emptycirc} & \midcell{\fullcirc} \tabularnewline \hline
14 & \apnfield{APN-OI-Replacement} & \valmid{<Attacker>.mnc<MNC>.mcc<MCC>.gprs} & \midcell{C1} & \midcell{\emptycirc} & \midcell{\fullcirc} \tabularnewline \hline
15 & \apnfield{Context-Identifier} & \valmid{Default PDN(1)} & \midcell{C1} & \midcell{\emptycirc} & \midcell{\fullcirc} \tabularnewline \hline
16 & \apnfield{Context-Identifier} & \valmid{Non-Default PDN (2)} & \midcell{C1} & \midcell{\emptycirc} & \midcell{\fullcirc} \tabularnewline \hline
17 & \apnfield{MIP6-Agent-Info.MIP6-Home-Agent-Address} & \valmid{1.1.1.1} & \midcell{C1} & \midcell{\emptycirc} & \midcell{\fullcirc} \tabularnewline \hline
18 & \apnfield{PDN-GW-Allocation-Type} & \valmid{STATIC} & \midcell{C1} & \midcell{\emptycirc} & \midcell{\fullcirc} \tabularnewline \hline
19 & \apnfield{Served-Party-IP-Address} & \valmid{1.1.1.1} & \midcell{C1} & \midcell{\emptycirc} & \midcell{\fullcirc} \tabularnewline \hline
20 & \apnfield{Service-Selection} & \valmid{default apn name} & \midcell{C1} & \midcell{\emptycirc} & \midcell{\fullcirc} \tabularnewline \hline
21 & \apnfield{Service-Selection} & \valmid{ims} & \midcell{C1} & \midcell{\emptycirc} & \midcell{\fullcirc} \tabularnewline \hline
22 & \apnfield{VPLMN-Dynamic-Address-Allowed} & \valmid{NOTALLOWED} & \midcell{C1} & \midcell{\emptycirc} & \midcell{\fullcirc} \tabularnewline \hline
23 & \apnfield{SIPTO-Local-Network-Permission} & \valmid{NOTALLOWED} & \midcell{C1} & \midcell{\emptycirc} & \midcell{\fullcirc} \tabularnewline \hline
24 & \apnfield{SIPTO-Permission} & \valmid{Null} & \midcell{C1} & \midcell{\emptycirc} & \midcell{\fullcirc} \tabularnewline \hline
25 & IDR: Subscription-Data.Regional-Subscription-Zone-Code & \val{0102} & C1 & \emptycirc & \fullcirc \tabularnewline \hline
26 & IDR: Subscription-Data.Subscriber-Status & \val{OPERATOR_DETERMINED_BARRING} & C1 & \emptycirc & \fullcirc \tabularnewline \hline
27 & IDR: Subscription-Data.Access-Restriction-Data & \val{WB-E-UTRAN Not Allowed} & C1 & \emptycirc & \fullcirc \tabularnewline \hline
28 & IDR: Subscription-Data.HPLMN-ODB & \val{1} & C1 & \emptycirc & \fullcirc \tabularnewline \hline
29 & IDR: Subscription-Data.Operator-Determined-Barring & \val{All Packet Oriented Services Barred} & C1 & \emptycirc & \fullcirc \tabularnewline \hline
30 & IDR: Subscription-Data.CSG-Subscription-Data.Expiration-Date & \val{2026-06-11 23:59:59} & C1 & \emptycirc & \fullcirc \tabularnewline \hline
31 & IDR: Subscription-Data.CSG-Subscription-Data.CSG-Id & \val{0x00000020} & C1 & \emptycirc & \fullcirc \tabularnewline \hline
32 & PUR: User-Name & \val{IMSI} & C2 & \emptycirc & \fullcirc \tabularnewline \hline
33 & \apnfield{PDN-Type} & \valmid{IPv6} & \midcell{C1} & \midcell{\emptycirc} & \midcell{\fullcirc} \tabularnewline \hline
34 & \apnfield{PDN-Type} & \valmid{IPv4v6} & \midcell{C2} & \midcell{\emptycirc} & \midcell{\fullcirc} \tabularnewline \hline
35 & \apnfield{EPS-Subscribed-QoS-Profile} & \valmid{QCI_9} & \midcell{C2} & \midcell{\emptycirc} & \midcell{\fullcirc} \tabularnewline \hline
36 & \apnfield{EPS-Subscribed-QoS-Profile.Allocation-Retention-Priority} & \valmid{15} & \midcell{C2} & \midcell{\emptycirc} & \midcell{\fullcirc} \tabularnewline \hline
37 & IDR: Subscription-Data.AMBR.Max-Requested-Bandwidth-DL & \val{1bps} & C2 & \emptycirc & \fullcirc \tabularnewline \hline
38 & IDR: Subscription-Data.AMBR.Max-Requested-Bandwidth-UL & \val{1bps} & C2 & \emptycirc & \fullcirc \tabularnewline \hline
39 & \apnfield{AMBR.Max-Requested-Bandwidth-DL} & \valmid{1bps} & \midcell{C2} & \midcell{\emptycirc} & \midcell{\fullcirc} \tabularnewline \hline
40 & \apnfield{AMBR.Max-Requested-Bandwidth-UL} & \valmid{1bps} & \midcell{C2} & \midcell{\emptycirc} & \midcell{\fullcirc} \tabularnewline \hline
41 & IDR: Subscription-Data.Subscribed-Periodic-RAU-TAU-Timer & \val{2s} & C4 & \emptycirc & \fullcirc \tabularnewline
\hline
\end{tabularx}
}
\end{subtable}

\end{table*}

\begin{table*}[!t]
\ContinuedFloat
\centering
\caption{Detailed results of the implementation-level assessment. Continued.}
\footnotesize
\setlength{\tabcolsep}{2pt}
\renewcommand{\arraystretch}{1.00}

\setcounter{subtable}{1}
\begin{subtable}[t]{\textwidth}
\centering
\vspace{-3pt}
\caption{Cx Interface}
\vspace{-3pt}
\label{table: affected-diameter-fields-cx}

{\renewcommand{\tabularxcolumn}[1]{m{#1}}
\begin{tabularx}{\textwidth}{|N|R|S|C|Y|Y|Y|}
\hline
\multirow{2}{*}{\textbf{PID}} &
\multirow{2}{*}{\textbf{Diameter Fields}} &
\multirow{2}{*}{\normalfont\textbf{Value}} &
\multirow{2}{*}{\textbf{CID}} &
\multicolumn{3}{|c|}{\affectedimpl} \tabularnewline
\cline{5-7}
&  &  &  & \textbf{O2} & \textbf{C1} & \textbf{C2} \tabularnewline
\hline
42 & RTR: User-Name & \val{000001234567890@ims.mnc000.mcc000.3gppnetwork.org} & C3 & \emptycirc & \fullcirc & \emptycirc \tabularnewline \hline
43 & RTR: Associated-Identities & \val{000001234567890@ims.mnc000.mcc000.3gppnetwork.org} & C1 & \fullcirc & \fullcirc & \emptycirc \tabularnewline \hline
44 & RTR: Public-Identity & \val{sip:000001234567890@ims.mnc000.mcc000.3gppnetwork.org} & C1 & \fullcirc & \fullcirc & \fullcirc \tabularnewline \hline
45 & SAR: Server-Name & \val{sip:scscf.ims.system.com} & C1 & \fullcirc & \fullcirc & \fullcirc \tabularnewline \hline
46 & SAR: Server-Assignment-Type & \val{REGISTRATION} & C1 & \fullcirc & \fullcirc & \emptycirc \tabularnewline \hline
47 & SAR: Server-Assignment-Type & \val{RE_REGISTRATION} & C1 & \fullcirc & \fullcirc & \emptycirc \tabularnewline \hline
48 & SAR: Server-Assignment-Type & \val{UNREGISTERED_USER} & C3 & \emptycirc & \fullcirc & \fullcirc \tabularnewline \hline
49 & SAR: Server-Assignment-Type & \val{USER_DEREGISTRATION} & C1 & \emptycirc & \fullcirc & \fullcirc \tabularnewline \hline
50 & SAR: Server-Assignment-Type & \val{ADMINISTRATIVE_DEREGISTRATION} & C1 & \emptycirc & \fullcirc & \fullcirc \tabularnewline \hline
51 & MAR: 3GPP-SIP-Number-Auth-Items & \val{Digest-AKAv1-MD5} & C3 & \fullcirc & \fullcirc & \fullcirc \tabularnewline \hline
52 & UAR: User-Authorization-Type & \val{REGISTRATION} & C3 & \fullcirc & \fullcirc & \fullcirc \tabularnewline \hline
53 & UAR: User-Authorization-Type & \val{REGISTRATION_AND_CAPABILITIES} & C3 & \emptycirc & \fullcirc & \emptycirc \tabularnewline \hline
54 & SAR: User-Data-Already-Available & \val{USER_DATA_NOT_AVAILABLE} & C3 & \fullcirc & \fullcirc & \fullcirc \tabularnewline \hline
55 & LIR: User-Authorization-Type & \val{REGISTRATION} & C3 & \fullcirc & \fullcirc & \fullcirc \tabularnewline \hline
56 & LIR: User-Authorization-Type & \val{REGISTRATION_AND_CAPABILITIES} & C3 & \emptycirc & \fullcirc & \emptycirc \tabularnewline \hline
57 & LIR: Originating-Request & \val{ORIGINATING} & C5 & \fullcirc & \fullcirc & \fullcirc \tabularnewline \hline
58 & PPR: 3GPP-SIP-Auth-Data-Item & \val{Any String} & C1 & \emptycirc & \fullcirc & \fullcirc \tabularnewline \hline
59 & PPR: Cx-User-Data & \val{Forged User Profile XML} & C1 & \emptycirc & \fullcirc & \fullcirc \tabularnewline \hline
60 & RTR: Deregistration-Reason.Reason-Code & \val{PERMANENT_TERMINATION} & C1 & \emptycirc & \fullcirc & \fullcirc \tabularnewline \hline
61 & RTR: Deregistration-Reason.Reason-Info & \val{Any String} & C1 & \emptycirc & \fullcirc & \emptycirc \tabularnewline \hline
62 & SAR: SCSCF-Restoration-Info.Path & \val{sip:term@192.168.5.103:5060} & C1 & \emptycirc & \fullcirc & \fullcirc \tabularnewline \hline
63 & SAR: SCSCF-Restoration-Info.Contact & \val{sip:75ba167d-7228-40d0-a1aa-54d9f5b9fbf1@1.1.1.1:11111} & C1 & \emptycirc & \fullcirc & \fullcirc \tabularnewline \hline
64 & SAR: Wildcarded-Public-Identity & \val{sip:000001234567890@ims.mnc000.mcc000.3gppnetwork.org} & C1 & \emptycirc & \fullcirc & \emptycirc \tabularnewline
\hline
\end{tabularx}
}
\end{subtable}

\end{table*}

\begin{table*}[!t]
\centering
\caption{Generalizability results across non-Diameter protocols.}
\label{table:generalizability}
\vspace{-3pt}
\fontsize{5.6pt}{6pt}\selectfont
\setlength{\tabcolsep}{2.5pt}
\renewcommand{\arraystretch}{1.00}

\resizebox{\textwidth}{!}{%
\begin{tabular}{
@{}r@{\hspace{8pt}}
l@{\hspace{12pt}}
l@{\hspace{10pt}}
c@{\hspace{7pt}}
c@{\hspace{7pt}}
c@{\hspace{7pt}}
r@{\hspace{2pt}}l@{\hspace{12pt}}
l@{\hspace{10pt}}
l@{}
}
\toprule
\textbf{No.} &
\textbf{Protocol} &
\textbf{Application Domain} &
\makecell[c]{\textbf{Analyzable}\\\textbf{Fields}} &
\makecell[c]{\textbf{Generated}\\\textbf{PoCs}} &
\makecell[c]{\textbf{New}\\\textbf{PoCs}} &
\multicolumn{2}{c@{\hspace{12pt}}}{\makecell[c]{\textbf{Confirmed New}\\\textbf{PoCs (\%)}}} &
\makecell[c]{\textbf{Tested}\\\textbf{Implementation}} &
\textbf{Version} \\
\midrule
\phantom{0}1  & SIP       & VoIP Communication          & \phantom{0}99 & 17 & 10 & \phantom{0}8 & (80.00\%)  & Kamailio        & 6.2.0-dev0 \\
\phantom{0}2  & BGP       & Internet Routing            & \phantom{0}78 & 22 & 15 & 12 & (80.00\%)  & FRRouting       & 10.3.1 \\
\phantom{0}3  & MQTT      & IoT Messaging               & \phantom{0}67 & 39 & 20 & 16 & (80.00\%)  & Mosquitto       & 2.1.2 \\
\phantom{0}4  & DHCP      & Network Configuration       & \phantom{0}27 & 24 & 10 & \phantom{0}8 & (80.00\%)  & Suricata/Snort3 & 3.11.1.0 \\
\phantom{0}5  & RADIUS    & Network Authentication      & \phantom{0}59 & 12 & \phantom{0}9 & \phantom{0}8 & (88.89\%)  & FreeRADIUS      & 3.2.8 \\
\phantom{0}6  & HTTP/2    & Web Infrastructure          & \phantom{0}34 & 24 & \phantom{0}2 & \phantom{0}2 & (100.00\%) & Nginx           & 1.27.x \\
\phantom{0}7  & DNS       & Internet Infrastructure     & \phantom{0}27 & 15 & \phantom{0}1 & \phantom{0}1 & (100.00\%) & DnsMasq         & 2.90 \\
\phantom{0}8  & SNMP      & Network Management          & \phantom{0}24 & \phantom{0}3 & \phantom{0}0 & \phantom{0}0 & (N/A)      & Net-SNMP        & 5.10.pre2 \\
\phantom{0}9  & CCSDS-SLE & Space Communication         & \phantom{0}46 & 26 & 25 & 22 & (88.00\%)  & CCSDS           & 1.1.0 \\
10            & OCI       & Container Runtime           & \phantom{0}54 & 23 & 11 & \phantom{0}7 & (63.64\%)  & Docker Desktop  & 4.58.0 \\
11            & Wayland   & Desktop Graphics            & \phantom{0}50 & 18 & 15 & 12 & (80.00\%)  & Weston/Sway     & 13.0.0 / 1.9 \\
12            & Modbus    & Industrial Control          & \phantom{0}39 & 15 & \phantom{0}5 & \phantom{0}3 & (60.00\%)  & OpenPLC         & v3 \\
13            & FUSE      & File-System Interface       & \phantom{0}34 & \phantom{0}9 & \phantom{0}5 & \phantom{0}3 & (60.00\%)  & S3fs-Fuse       & 1.97 \\
14            & RTSP      & Multimedia Streaming        & \phantom{0}73 & \phantom{0}5 & \phantom{0}3 & \phantom{0}2 & (66.67\%)  & live555         & 2026-03-23 \\
15            & AMQP      & Enterprise Messaging        & \phantom{0}69 & 11 & \phantom{0}9 & \phantom{0}7 & (77.78\%)  & ActiveMQ        & 2.37.0 \\
16            & D-Bus     & Inter-Process Communication & \phantom{0}77 & 16 & 10 & \phantom{0}7 & (70.00\%)  & Dbus-Daemon     & 1.12.20 \\
\midrule
\textbf{} & \textbf{} & \textbf{Total}
& \textbf{857}
& \textbf{279}
& \textbf{150}
& \textbf{118}
& \textbf{(78.67\%)}
& \textbf{} & \textbf{} \\
\bottomrule
\end{tabular}%
}
\vspace{-0.6em}
\end{table*}

\begin{table*}[!t]
\centering
\caption{Comparison with DiaFuzzer-based fuzzing.}
\label{table: fuzzing comparison}
\vspace{-3pt}

\footnotesize{}
\setlength{\tabcolsep}{4pt}
\renewcommand{\arraystretch}{1.00}

\newlength{\effectnumwidth}
\newlength{\effectpctwidth}
\settowidth{\effectnumwidth}{26}
\settowidth{\effectpctwidth}{(100.00\%)}

\newcommand{\effectcell}[2]{%
  \makebox[\effectnumwidth][r]{#1}%
  \hspace{1pt}%
  \makebox[\effectpctwidth][l]{(#2\%)}%
}

\begin{tabular*}{\textwidth}{
@{\hspace{\tabcolsep}\extracolsep{\fill}}
l
c
r
r
r
r
@{\hspace{\tabcolsep}}
}
\toprule

\multirow{2}{*}{\textbf{Implementation}}
& \multirow{2}{*}{\textbf{Interface}}
& \multicolumn{2}{c}{\textbf{\name{}}}
& \multicolumn{2}{c}{\textbf{DiaFuzzer-Based Fuzzing}} \\

\cmidrule(lr){3-4}
\cmidrule(lr){5-6}

&
& \multicolumn{1}{c}{\textbf{Confirmed Fields}}
& \multicolumn{1}{c}{\textbf{Confirmed PoCs}}
& \multicolumn{1}{c}{\textbf{PoC Manipulations Generated}}
& \multicolumn{1}{c}{\textbf{PoC Effects Detected}} \\

\midrule

Open5GS
& S6a
& 4
& 7
& 7
& \effectcell{7}{100.00} \\

Kamailio
& Cx
& 9
& 10
& 10
& \effectcell{7}{70.00} \\

Commercial \textit{A}
& S6a
& 35
& 40
& 40
& \effectcell{12}{30.00} \\

Commercial \textit{A}
& Cx
& 16
& 23
& 23
& \effectcell{12}{52.17} \\

Commercial \textit{B}
& Cx
& 15
& 15
& 15
& \effectcell{7}{46.67} \\

\midrule

\textbf{Overall}
& --
& \textbf{51}
& \textbf{64}
& \textbf{64}
& \textbf{\effectcell{26}{40.63}} \\

\bottomrule
\end{tabular*}

\vspace{6pt}

{\small
\parbox{\textwidth}{
\textit{Note:} A field may correspond to multiple PoCs when different
field values cause distinct security effects. PoC Manipulations Generated
counts \name{}-confirmed PoCs for which DiaFuzzer-based fuzzing generates
the corresponding field-value manipulation, while PoC Effects Detected
counts PoCs whose security effects are also automatically detected by the
fuzzer. The Overall row reports deduplicated counts across implementations.
}}

\end{table*}

\section{Ethics Considerations}
We conducted all experiments in controlled test environments. For Open5GS and Kamailio, we deployed local testbeds inside a Faraday cage, which were not connected to production mobile networks, real subscriber databases, or public operator infrastructure. This setup prevented our tests from affecting nearby devices or normal users. For commercial systems, we obtained permission from the vendors and tested only on systems isolated from their real deployments.

For the non-Diameter protocols, we built local test environments using open-source implementations and evaluated the generated PoCs only in these environments. All test inputs and failures were confined to our local machines or virtual machines.

We have reported a subset of the confirmed vulnerabilities to the affected vendors, open-source projects, CVE, and CNVD. We are preparing the remaining reports for phased responsible disclosure, which will be completed for all confirmed vulnerabilities no later than the rebuttal deadline, and will not release PoCs for unfixed vulnerabilities.

The main way for us and future users to limit the impact of LENS is to run it only in controlled environments and to report confirmed vulnerabilities to the affected developers in a timely manner. Because LENS is designed to find vulnerabilities, it could still be misused to test systems without permission or to disrupt vulnerable deployments. However, we believe this remaining risk is outweighed by the security benefit of reporting these flaws, since the results help affected systems fix bugs before they are exploited in real deployments.

\newpage 


\section{Meta-Review}
The following meta-review was prepared by the program committee for the 2027 IEEE Symposium on Security and Privacy (S\&P) as part of the review process as detailed in the call for papers.

\subsection{Summary}
This paper presents LENS, an LLM-based framework for field-level exploitability analysis of Diameter. LENS derives an analysis procedure from expert examples, collects protocol evidence, and uses it to assess fields and construct PoCs. The evaluation covers Open5GS, Kamailio, and two commercial implementations.

\subsection{Scientific Contributions}
\begin{itemize}
\item Creates a New Tool to Enable Future Science.
\item Identifies an Impactful Vulnerability.
\item Provides a Valuable Step Forward in an Established Field.
\end{itemize}

\subsection{Reasons for Acceptance}
\begin{enumerate}
\item LENS provides a practical way to identify semantic security effects that conventional fuzzing oracles can miss. The paper also reports concrete findings and working PoCs across open-source and commercial implementations.
\item The authors tested four LLMs with five runs each and manually audited all 223 fields initially reported by LENS. After removing 19 unsupported results, 51 fields were confirmed in at least one implementation and 153 were validated at the protocol level. Grouping these fields into 49 message-level scenarios also makes the scope of the results clearer.
\end{enumerate}

\subsection{Noteworthy Concerns} 
\begin{enumerate} 
\item The threat model requires the attacker to operate as an accepted Diameter peer. Deployment assumptions are not clearly stated, and protocol-defined peer capabilities are not clearly distinguished from implementation vulnerabilities.
\item The reliability benchmark contains only 40 author-labeled fields, which limits reliability and cross-protocol generalization.
\item DiaFuzzer is given the same mutations as LENS. This comparison therefore evaluates effect detection, not end-to-end vulnerability discovery.
\end{enumerate}

\end{document}